\documentclass[
 aps, pra,
 amsmath,amssymb,
 11pt,
 final,
tightenlines,
 twoside,
 twocolumn,
 nofloats,
nofootinbib,
 superscriptaddress,
showkeys,
showkeywords,
 ]
{revtex4-2}

\usepackage[T2A]{fontenc}
\usepackage[utf8x]{inputenc}
\usepackage[russian,english]{babel}
\usepackage{graphicx}% Include figure files
\usepackage{dcolumn}% Align table columns on decimal point
\usepackage{bm}% bold math
\usepackage{url}

\input{maik.rty}

\setcitestyle{authoryear,round}
\def\saoname{Special Astrophysical Observatory,  Russian Academy of Sciences,
              Nizhnii Arkhyz, 369167 Russia}

\def\squareforqed{\hbox{\rlap{$\sqcap$}$\sqcup$}}

\def\sq{\ifmmode\squareforqed\else{\unskip\nobreak\hfil
\penalty50\hskip1em\null\nobreak\hfil\squareforqed
\parfillskip=0pt\finalhyphendemerits=0\endgraf}\fi}

\def\arcmin{\hbox{$^\prime$}}

\def\arcsec{\hbox{$^{\prime\prime}$}}

\def\utw{\smash{\rlap{\lower5pt\hbox{$\sim$}}}}

\def\udtw{\smash{\rlap{\lower6pt\hbox{$\approx$}}}}

\def\diameter{{\ifmmode\mathchoice
{\ooalign{\hfil\hbox{$\displaystyle/$}\hfil\crcr
{\hbox{$\displaystyle\mathchar"20D$}}}}
{\ooalign{\hfil\hbox{$\textstyle/$}\hfil\crcr
{\hbox{$\textstyle\mathchar"20D$}}}}
{\ooalign{\hfil\hbox{$\scriptstyle/$}\hfil\crcr
{\hbox{$\scriptstyle\mathchar"20D$}}}}
{\ooalign{\hfil\hbox{$\scriptscriptstyle/$}\hfil\crcr
{\hbox{$\scriptscriptstyle\mathchar"20D$}}}}
\else{\ooalign{\hfil/\hfil\crcr\mathhexbox20D}}%
\fi}}

\newcommand{\aap}{Astron. and Astrophys. }
\newcommand{\aj}{Astron.~J. }
\newcommand{\apss}{Astrophys. and Space Sci. }
\newcommand{\araa}{Annual Rev. Astron. Astrophys. }
\newcommand{\mnras}{Monthly Notices Royal Astron. Soc. }
\newcommand{\apjl}{Astrophys.~J. Letters }

\usepackage{xcolor,soul}

\newcommand{\be}{\begin{equation}} 
\newcommand{\ee}{\end{equation}}
\newcommand{\obj}{Ho\,II~X-1}

\newcommand{\astrosat}{AstroSat}

\begin{document}
\selectlanguage{english}
%\keywords{ULX, UVIT, SCAD }

\title{Simultaneous X-ray/UV observations of ultraluminous X-ray source Holmberg\,II~X-1 with Indian space mission AstroSat}

\author{\firstname{A.}~\surname{Vinokurov}}
 \email{vinokurov@sao.ru}
 \affiliation{\saoname}

\author{\firstname{K.}~\surname{Atapin}}
 \affiliation{Sternberg Astronomical Institute, Moscow State University, Universitetsky pr., 13, Moscow, 119991, Russia}
 \affiliation{\saoname}

\author{\firstname{O.~P.}~\surname{Bordoloi}}
 \affiliation{Tezpur University, Tezpur, Assam 784028, India}

\author{\firstname{A.}~\surname{Sarkisyan}}
 \affiliation{\saoname}

\author{\firstname{U.}~\surname{Kashyap}}
 \affiliation{Indian Institute of Technology Indore, Indore 453552, India}

\author{\firstname{M.}~\surname{Chakraborty}}
 \affiliation{Indian Institute of Technology Indore, Indore 453552, India}

\author{\firstname{P.~T.}~\surname{Rahna}}
 \affiliation{CAS Key Laboratory for Research in Galaxies and Cosmology, Shanghai Astronomical Observatory, Shanghai 200030, China}

\author{\firstname{A.}~\surname{Kostenkov}}
 \affiliation{\saoname}

\author{\firstname{Y.}~\surname{Solovyeva}}
 \affiliation{\saoname}

\author{\firstname{S.}~\surname{Fabrika}}
 \affiliation{\saoname}

\author{\firstname{M.}~\surname{Safonova}}
 \affiliation{Indian Institute of Astrophysics, Bengaluru 560034, India}

\author{\firstname{R.}~\surname{Gogoi}}
 \affiliation{Tezpur University, Tezpur, Assam 784028, India}

\author{\firstname{F.}~\surname{Sutaria}}
 \affiliation{Indian Institute of Astrophysics, Bengaluru 560034, India}

\author{\firstname{J.}~\surname{Murthy}}
 \affiliation{Indian Institute of Astrophysics, Bengaluru 560034, India}

\begin{abstract}
%Despite the intensive studies of the ultraluminous X-ray sources (ULXs) in both X-rays and in optical, and many indirect arguments (relation to the youngest population, non-standard accretion disks), there are no strong and obvious evidences to distinguish among two competitive models of the ULXs: whether they contain stellar-mass or intermediate-mass black holes (IMBHs). What we know exactly is that the ULXs are close binary systems with massive donors (UV--optical spectral energy distributions hint at the two-component spectra: disk (or wind) and a donor). Recent data show that both UV and optical emissions may be reprocessed in strong heating by X-rays. To test the components of the ULX-binaries and the UV responses to the X-ray variability, we obtained simultaneous UV and X-ray observations of the highly variable ULX Holmberg~II X-1 with \astrosat --- Indian multiwavelength space satellite. Our observations with SXT and UVIT payloads onboard show possible correlation between UV and X-ray fluxes, which corresponds to a binary with the stellar-mass black hole and the supercritical accretion disk, rather than a binary with the IMBH.    

We present the results of 8 epochs of simultaneous UV and X-ray observations of the highly variable ultraluminous X-ray source (ULX) Holmberg~II X-1 with \astrosat --- Indian multiwavelength space satellite. During the entire observation period from late 2016 to early 2020, Holmberg~II X-1 showed a moderate X-ray luminosity of $\approx8\times10^{39}$ erg~s$^{-1}$ and a hard power-law spectrum with $\Gamma \lesssim 1.9$. Due to low variability of the object in X-rays (by a factor 1.5) and insignificant variability in the UV range (upper limit $\approx25$\%) we could not find reliable correlation between flux changes in these ranges. Inside each particular observation, the X-ray variability amplitude is higher, it reaches a factor of 2-3 respect to the mean level at the time scales of $\sim10$ ks or even shorter. We discussed our results in terms of three models of a heated donor star, a heated disk and a heated wind, and estimated the lower limit to the variability which would allow to reject at least part of them. %Despite the low variability of the object by a factor less than 2, we find a possible correlation between X-ray and optical radiation with a coefficient of up to $0.87 \pm 0.20$ ($pval = 0.005$), which nevertheless strongly depends on the choice of aperture for measuring the X-ray flux. Correlated flux changes with comparable amplitudes may indicate a relatively low contribution of the donor to UV radiation, which excludes supergiants of early spectral classes.
% На основе простых соотношений мы показываем, что если донорами некоторых ярких в оптическом диапазоне ULXs, таких как Holmberg\,II~X-1, являются гиганты или сверхгиганты ранних спектральных классов, то высокая переменность таких объектов в УФ диапазоне должна быть обусловлена изменениями светимости аккреционного диска, но не прогревом донора.
%may be evidence of a relatively weak contribution of the donor to UV radiation, which excludes supergiants of early spectral classes.% as candidates.
\end{abstract}

\maketitle
\section{Introduction}
By definition, ultraluminous X-ray sources (ULXs) are non-nuclear point-like extragalactic sources with apparent luminosities exceeding the Eddington limit for typical black holes in our Galaxy. According to modern view, the most of them are binary systems with stellar-mass black holes and neutron stars accreting at super-Eddington rates \citep{Kaaret2017review,Fabrika2021review}. Presence of neutron stars among ULXs became evident after the discovery of a coherent pulsation of X-ray emission in the well-known ULX M82 X-2 \citep{Bachetti2014Nat}. Now, about a dozen ultraluminous pulsars (ULXP) are known.

A key feature of super-Eddington accretion is strong outflows originating from the innermost parts of the supercritical accretion disk \citep{Poutanen2007}. Spectroscopic evidence of such outflows has been revealed in both X-ray and optical ranges, however, properties of these outflows seem to be different between the ranges. The X-ray spectral lines produced by the outflowing matter are highly blueshifted indicating that the gas moves with velocities of about $0.1c$ (so-called ultra-fast outflows, \citealt{Pinto2016Nat}). In the optical range, the outflow looks similar to stellar winds and has velocities of $500-1500$ km/s \citep{Fabrika2015}. This wind is believed to be optically thick and has to form extended photosphere around the supercritical accretion disk. Using the methods developed for the modeling of optical spectra of Wolf-Rayet and LBV stars, it has been shown that the mass loss rate in the ULX wind is about $10^{-6}-10^{-5.5}$ M$_\odot$/y \citep{Kostenkov2020a}.

To provide required accretion rates the donor star must be in a close orbit and, highly likely, fills its Roche-lobe (though the first discovered donor of ULX was turned out to be a Wolf-Rayet star, M101 ULX-1 \citealt{Liu2013NaturM101ULX1}). The donors are known only for a handful of ULXs and in most cases they are blue or red supergiants \citep{Motch2014NaturP13, Heida2015NGC253,Heida2016fiveULXs,Heida2019NGC300,Lopez2020}. Nevertheless, the significant part of optical emission is believed to originate not from the donor but from the wind photosphere of the supercritical disk (\citealt{Fabrika2021review} and reference therein). It was shown that the brightest in the optical range ULXs have blue, power-law spectral energy distributions (SEDs) with a maximum in far-UV band beyond the observable range of wavelength \citep{Tao2012HoIIoptical,Griese2012NGC5408X1, Vinokurov2013}. The less bright sources show SEDs consistent with spectra of A-G class stars \citep{Avdan2016,Avdan2019}. This may suggest that the spectra of the brightest counterparts are fully dominated by optical emission of the hot supercritical accretion disc, but as the disc contribution decreases, the donor emission become more prominent \citep{Vinokurov2018}.

%{\bf In this regard, the joint analysis of the optical and X-ray emission of ULXs is an urgent task. Both the X-ray and optical emission are powered by the same source - the central machine of the super In this regard, the joint analysis of the optical and X-ray emission of ULXs is an urgent task. Both the X-ray and optical emission are powered by the same source - the central machine of the supercritical disk and have to be related to each other \cite{Gierlinski, Fabrika}. Therefore such studies may shed light on details of wind formation and outflow geometry. Some advances in this field have al

%Both the X-ray and optical emission of a supercritical disk are powered by the same source, the central machine of the supercritical disk, and have to be related to each other \cite{Gierlinski, Fabrika}. 
It is assumed that the optical radiation of the supercritical disk is a product of the re-processing of X-ray quanta in outer parts of the accretion disk or wind \citep{Gierlinski2009,Fabrika2015}. Moreover, depending on the physical characteristics of the region of re-emission of hard quanta and the inclination angle of the accretion disk to the line of sight, the ratio between the fluxes in these two ranges can differ greatly. %, that implies certain dependencies between the fluxes in these two ranges}
In this regard, the joint analysis of the optical and X-ray emission of ULXs is an urgent task. Such studies may shed light on details of wind formation and outflow geometry. Some advances in this field have already been made. \citet{Griese2012NGC5408X1} obtained simultaneous observations of NGC\,5408~X-1 with the Chandra X-ray Observatory and the Hubble Space Telescope, and constructed the source SED covering wavelengths from X-rays to near-IR. However, the source showed weak variability during those observations, which did not allow the authors to test for the presence of a correlation between these bands. 

%{\bf A reliable correlation between the UV-optical and X-ray fluxes was reported for a few transient ULXs: in the literature we have found such evidence for UGC\,6456 ULX \citep{Vinokurov2020}, NGC\,7793 P13 in some orbital periods when phase difference arising due to presence of super-orbital period between X-ray and UV-optical light curve becomes 0 \citep{FurstWaltonHeida2018}, and NGC\,300 ULX-1 during its famous outburst in 2010 \citep{VillarBerger2016}. Notes that the last two objects are pulsating ULXs with neutron stars. }

%{\bf A reliable correlation between the UV-optical and X-ray fluxes has been reported for a few transient ULXs. For example, in the case of UGC\,6456 ULX, not only the correlation between the X-ray and optical fluxes was revealed, but also indirectly between the X-ray flux and the HeII\,$\lambda4686$ emission line flux \citep{Vinokurov2020}, however, the physical reasons for the object variability at different wavelengths are not considered due to the small number of observations. }

A reliable correlation between the UV-optical and X-ray fluxes has been reported for a few transient ULXs. For example, in the case of UGC\,6456 ULX, the X-rays were found to be positively correlated not only with broadband optical emission but also, indirectly, with the flux in the HeII\,$\lambda4686$ emission line \citep{Vinokurov2020}, however, the physical mechanisms of the observed variability at different wavelengths were not considered due to the small number of observations.

%Here we demonstrate the capabilities of studying ULX with \astrosat\footnote{\astrosat is an Indian first dedicated astronomy satellite. Detail information about the \astrosat mission and all instruments aboard is available on the \astrosat Science Support Cell webpage (http://astrosat-ssc.iucaa.in)} using the Holmberg\,II X-1 (herafter \obj) as an example. 
Here we present the results of multiple simultaneous UV--X-ray observations of Holmberg\,II X-1 (herafter \obj) with AstroSat\footnote{Detail information about the \astrosat\ mission and all instruments aboard is available on the \astrosat\ Science Support Cell webpage (http://astrosat-ssc.iucaa.in)}, an Indian first dedicated astronomy satellite aimed at studying celestial sources in X-ray, optical and UV spectral bands simultaneously \citep{Singh2014astrosat}. 
We measured the object fluxes in the X-ray and UV ranges and tried to search for a correlation between them. Also we discussed several models that could produce such a correlation.

\section{AstroSat capabilities for ULX observations}
\astrosat\ was launched on 2015 September 8 into a 650 km orbit from Satish Dhawan Space Centre, Sriharikota, India. The satellite is equipped with three X-ray instruments: Soft X-ray Telescope (SXT), Large Area X-ray Proportional Counters (LAXPCs) and Cadmium Zinc Telluride Imager (CZTI), as well as Ultra Violet Imaging Telescope (UVIT), providing coverage of $0.3-100$~keV in X-rays and of $1300-5500$~\AA\ in the UV/optical range.

\subsection{Ultra Violet Imaging Telescope}
UVIT constitutes two co-aligned Ritchey-Chretien 37.5-cm telescopes, one of which feeds a far-ultraviolet detector (FUV: $1300 - 1800$ \AA), with the other feeding two detectors through a beam splitter in the near-ultraviolet (NUV: $1800 - 3000$ \AA) and visible (VIS: $3200 - 5500$ \AA), providing imaging with a $28'$ field of view (FOV) and pixel scale of $\approx0.41\arcsec$. Spatial resolution achievable in the UV bands can be 1.2\arcsec~--1.6\arcsec, and is mainly determined by the quality of the correction for the spacecraft motion at the data prepossessing stage (see Sec.\,3.1), which in turn depends on the presence of bright sources in the field. Each of two UV channels has a filter wheel with 8 slots. %\footnote{https://uvit.iiap.res.in/Instrument/Filters}, %properties of the filters used in this work is listed in Table~\ref{table:filters}.
In the broad-band filters, the photometric sensitivity of the instrument is about 35\% that of GALEX \citep{Rahna2017}.

The photons can be detected either in photon-counting (PC) or integration mode by $512\times 512$ CMOS detectors (identical for each channel) with image intensifiers consisting of photocathode and microchannel plates. In the PC mode, the detector reads a frame 29 times a second with an exposure time of 35 ms per frame. The VIS channel is used only to correct for the spacecraft motion, it always works in integration mode. % with an exposure time \chk{xxx}.

The NUV channel went out of order on 2018 March 20 and has never been used after this date.

\subsection{Soft X-ray Telescope}
The SXT is capable of providing X-ray images and spectra in the energy range $0.3-8$~keV.
It employs focusing optics of the Wolter-I type and a $600\times 600$ pixel CCD camera similar to those used in the XMM and Swift missions. The SXT has an effective area of 200 cm$^{2}$ at 1.5 keV and a focal length of 2 meters.
The size of the point spread function (PSF) is 3\arcmin--4\arcmin, the half-power diameter (HPD) is 10\arcmin, the telescope field of view is $\approx 40$\arcmin. In the case of observations of faint sources, the detector works in photon counting mode with time resolution of 2.4 s. 

%SXT has an effective area of 200 cm$^{2}$ at 1.5 keV and a focal length of 2 meters. The size of the point spread function (PSF) is $3^{\prime}-4^{\prime}$ and telescope field of view is $\approx 40^{\prime}$. The on-axis FWHM and half-power diameter (HPD) of the PSF in the focal plane are $2'^{\prime}$ and $10^{\prime}$ respectively. The time resolution of SXT is 2.4 s. The good sensitivity, spatial and spectral resolution makes SXT favourable for spectral analysis as well as variability observations in the soft X-ray regime. 

%\subsection{Large Area X-ray Proportional Counters}
\subsection{Hard X-ray detectors}
Two other detectors: a set of xenon-filled proportional counters~--- the LAXPC and a pixellated cadmium-zinc-telluride array with a coded aperture mask~--- the CZTI, cover energies from about 10 to 150~keV. Despite the large effective area (LAXPC has $\sim6000$\,cm$^2$), these detectors have poor spatial resolution and, therefore, high background. This makes them inefficient for observing extragalactic sourсes like ULXs. Moreover, observations of ULXs with NuStar have shown a sharp drop of their X-ray flux above 15--20\,keV (a set of ULXs~--- \citealt{Walton2018puls}, and \obj\ in particular~--- \citealt{Walton2015}).
Inspection of the LAXPC data has shown that the signal remaining after background subtraction is statistically insignificant in all the observations of \obj. Therefore, we do not use the LAXPC and CZTI data in our analysis below.

Being in a low near-equatorial orbit, \astrosat\ periodically (in about 2/3 of orbits) passes the South Atlantic Anomaly (SAA). This region is characterized by high density of charged particles which is permanently measured by the Charged Particle Monitor. The monitor may command the scientific instruments to go into sleeping mode until the particle level become safe for them.

\section{Target, observations and data reduction}
\obj\ belongs to the most luminous known ULXs ($L_X > 10^{40}$ erg\,cm$^{-2}$\,s$^{-1}$, \citealt{Grise2010HoIIx1}). It is located in the Holmberg\,II irregular gas-rich dwarf galaxy, which is a part of the M\,81--NGC\,2403 group of galaxies. The distance to the galaxy is 3.39\,Mpc \citep{Karachentsev2002}. The galaxy has a low oxygen abundance, corresponding to a metallicity of $\approx0.1$\,Z$_\odot$ \citep{Pilyugin2014}.
%{Its coordinates are R.A.\,=\,08:19:28.99, Dec.\,=\,+70:42:19.4.}
We have chosen \obj\ as a target for \astrosat\ due to its high variability in X-rays, exceeding an order of magnitude at timescales from a few days to months \citep{Grise2010HoIIx1}, and also due to its high brightness in the UV range \citep{Vinokurov2013}. 
Fitting various stellar spectral models to the combined non-simultaneous data set from UV to near IR bands,  \cite{Tao2012HoIIoptical} has shown that both low-metallicity late O-star spectra and the irradiated disk model are equally acceptable. The model of a supercritical accretion disk with outflow of matter in the form of a wind also gives a comparable result \citep{Vinokurov2013,Vinokurov2017}. Studies of the object in the IR range allow to suspect that the donor in this binary may be a supergiant B[e] star \citep{Lau2017}. \obj\ had to be ejected from a young star cluster; a possible kinematic evidence of this event has been found by \cite{Egorov2017}. The ULX is surrounded by bright compact nebula \citep{Pakull2002}. %, and the size of its most bright part in high spatial resolution Hubble Space Telescope (HST) images is $\approx2.5$\arcsec. 
Since the nebula contribution in the UV range is much lower than in the optical and near-IR ones, the short wavelengths are more preferable for studying optical counterparts of  ULXs.

%The choice of the target for observations with AstroSat is due to its high X-ray variability, exceeding an order of magnitude at times from a few days to months \citep{Grise2010HoIIx1}, and due to high X-1 brightness in UV range \citep{Vinokurov2013}. Fitting various stellar spectral models to the combined, non-simultaneous data set from UV to near IR bands, \cite{Tao2012HoIIoptical} found that low-metallicity late O-star spectra or irradiated disk model provide acceptable fits. Holmberg X-1 is surrounded by bright compact nebula \citep{Pakull2002}, the size of its most bright part in high spatial resolution Hubble Space Telescope (HST) images is $\approx2.5$\arcsec. Since the nebula contribution is much lower in UV than in the optical and near-IR ranges, short wavelengths are most preferable for studying the optical counterpart of the ULX.

We observed \obj\ with \astrosat\ on ten epochs, with the dates and other details provided in Table~\ref{tab:observations}. All the instruments were co-aligned and generated data simultaneously with UVIT as a primary instrument. The filters used in UV observations and their properties are shown in Table \ref{table:filters}. 
%\hide{Inspection of the LAXPC data has shown that \obj\ is inaccessible for this instrument. Thus, below we analyze the observations only from UVIT and SXT.} 

\begin{figure*}
\centering
\includegraphics[scale=0.61]{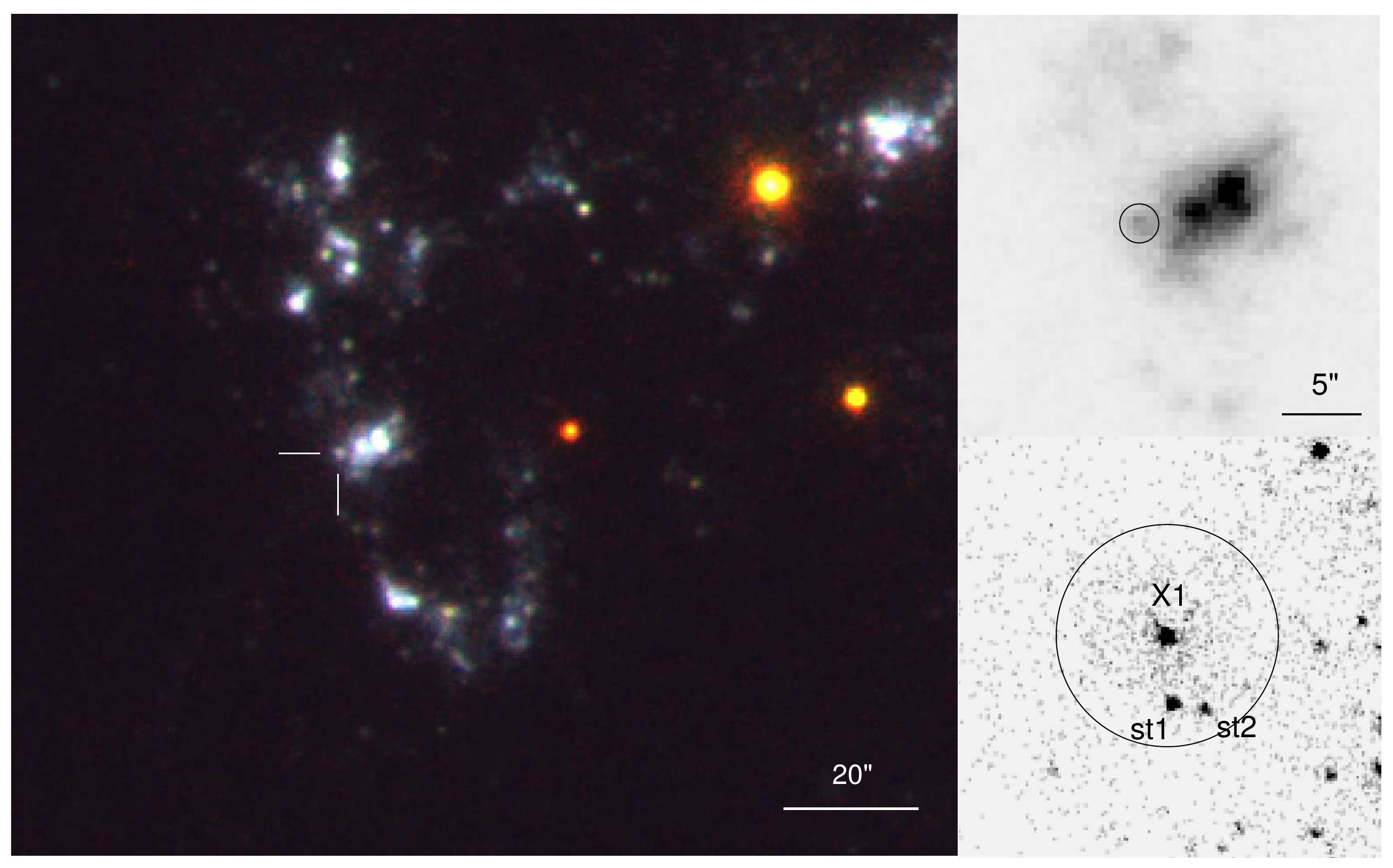}%\hspace{5mm}
\caption{{\it Left:} true color image of the \obj\ surroundings with blue, green and red channels corresponding to the total images in the F154W, N245M and N279N UVIT filters, respectively. {\it Right top:} local area of \obj\ in the filter F148W (from observation {\#}1, Table\,\ref{tab:observations}). The circle denotes the 1.23\arcsec\ aperture used in photometry of all the observations. {\it Right bottom:} HST ACS/SBC/F165LP image demonstrating that the aperture actually captures the ULX nebula and two neighboring start of comparable brightness unresolvable by \astrosat.}
\label{fig:fig1}
\end{figure*}

\begin{table*}
\newcommand{\twodates}[2]{\begin{tabular}{c}#1\\#2\end{tabular}}
\newcommand{\twodatesmultirow}[3]{\multirow{#1}{*}{\twodates{#2}{#3}}}
\caption{Observation dates, exposures and measured source fluxes. The X-ray fluxes (0.7-7.0\,keV) are for two extraction methods (I/II, Sec.\,3.2) in units of $10^{-12}$ erg~s$^{-1}$~cm$^{-2}$. The UV flux densities are background subtracted but accounted for flux contribution from neighboring sources (Sec.\,3.1.1), in units of $10^{-16}$ erg~s$^{-1}$~cm$^{-2}$~\AA$^{-1}$.}
\label{tab:observations}
\begin{tabular}{lcccccc}
\hline\hline  
N. & ObsID & Start time &  Channel & Flux & $T_{\rm exp}$,\\
 & & Stop time  & (filter) &    &  ks \\ \hline
 & & \twodatesmultirow{3}{2016-09-29 11:05:10}{2016-09-30 18:00:59}  & SXT & $3.90 \pm 0.22/5.6 \pm 0.4$ & 32.8 \\
1 & G05\_204T01\_9000000688~~ & & F148W & $3.59 \pm\ 0.11$ & 11.2 \\
 & & & N279N & $0.85 \pm\ 0.07$ & 10.6 \\ \hline
 
& & \twodatesmultirow{3}{2016-11-21 10:05:38}{2016-11-21 19:02:48} & SXT & $4.2 \pm 0.4/6.2 \pm 0.6$ & 11.8 \\
2 & A02\_046T01\_9000000814~~ & & F154W & $3.60 \pm\ 0.12$ & 9.0 \\
& & & N245M & $1.20 \pm\ 0.03$ & 8.2 \\ \hline

& & \twodatesmultirow{3}{2016-12-08 16:09:32}{2016-12-09 09:25:44} & SXT & $5.6 \pm 0.4/6.3 \pm 0.7$ & 13.0 \\
3 & A02\_046T01\_9000000864~~ & & F154W & $3.40 \pm\ 0.12$ & 9.1 \\
& & & N245M & $1.11 \pm\ 0.04$ & 8.7 \\ \hline

4 & A07\_054T01\_9000003286~~ & \twodates{2019-11-06 13:40:15}{2019-11-07 16:08:24} & SXT & $4.6 \pm 0.3/7.0 \pm 0.6$ & 18.6 \\ \hline
 
5 & A07\_054T01\_9000003348~~ & \twodates{2019-11-30 20:04:51}{2019-12-01 18:19:17} & SXT & $5.1 \pm 0.3/6.4 \pm 0.5$ & 22.8 \\ \hline
  
\multirow{2}{*}{6} & \multirow{2}{*}{A07\_054T01\_9000003370~~} & 2019-12-16 12:10:41 & SXT & $6.4 \pm 0.3/7.7 \pm 0.5$ & 24.0 \\
& & 2019-12-17 13:01:34 & F148W & $3.48 \pm\ 0.08$ & 17.4 \\ \hline

\multirow{2}{*}{7} & \multirow{2}{*}{A07\_054T01\_9000003378~~} & 2019-12-19 23:11:21 & SXT & $5.5 \pm 0.3/6.9 \pm 0.5$ & 26.8 \\
& & 2019-12-21 07:44:29 & F148W & $3.55 \pm\ 0.09$ & 17.2 \\ \hline

\multirow{2}{*}{8} & \multirow{2}{*}{A07\_054T01\_9000003406~~} & 2020-01-02 00:36:10 & SXT & $5.8 \pm 0.3/6.9 \pm 0.5$ & 23.0 \\
& & 2020-01-02 22:55:01 & F148W & $3.59 \pm\ 0.09$ & 16.8 \\ \hline

\multirow{2}{*}{9} & \multirow{2}{*}{A07\_054T01\_9000003486~~} & 2020-02-07 17:18:19 & SXT & $5.2 \pm 0.3/7.0 \pm 0.5$ & 23.1 \\
& & 2020-02-08 18:48:29 & F148W & $3.45 \pm\ 0.08$ & 16.6 \\ \hline

\multirow{2}{*}{10} & \multirow{2}{*}{A07\_054T01\_9000003504~~} & 2020-02-15 08:58:50 & SXT & $5.4 \pm 0.3/6.5 \pm 0.5$ & 23.2 \\
& & 2020-02-16 08:34:22 & F148W & $3.40 \pm\ 0.08$ & 17.1 \\ \hline
\end{tabular}\\
\end{table*}

\begin{table*}
\centering
\caption{Properties of the UVIT filters used in the observations. The effective bandwidth is the integral of the normalized effective area. 
The `mean $\lambda$' is the central source-independent wavelength. The conversion factors are taken from the \astrosat\ web page \texttt{https://uvit.iiap.res.in/Instrument/Filters}.}
\label{table:filters}
\begin{minipage}{14.7cm}
\begin{tabular}{cclcccc}
\hline 
Channel & ~~Slot~~ & Filter~~ & ~~~Bandpass~~ & ~~Effective bandwidth~~ & ~~Mean $\lambda$~~ & Conversion factor \\
 &  &  & ~~~~~~(nm) &  ~$\Delta\lambda$ (nm) &  (nm) & (erg/cm$^2$/\AA/cnt)\\     
\hline
\multicolumn{1}{c}{\multirow{2}{*}{FUV}} & 1 & F148W  & ~~~$125 - 179$ & 44.1  &  148.1  & 3.127e-15 \\
& 2 & F154W & ~~~$133 - 183$ & 37.8  &  154.1  & 3.593e-15 \\
\hline
\multirow{2}{*}{NUV}  & 3 & N245M  &  ~~~$220 - 265$ &  28.17     &    244.7   &    0.725e-15   \\
& 6 & N279N          &  ~~~$273 - 288$ & 8.95       &   279.2   &  3.500e-15\\
\hline
\end{tabular}\\\vspace{1mm}
\raggedright
\end{minipage}
\end{table*}

%All \astrosat payloads typically operate in photon counting mode, except the VIS channel on UVIT, which operates in integrating mode. Data from the \astrosat is transmitted to the \astrosat Data Centre at the Indian Space Science Data Centre (ISSDC), where they are separated by instrument and written into Level 1 data files, which are released to the PI of the proposal. When configuring for the observations, the PI specifies the primary instrument and secondary instruments. 

\subsection{UV observations} 
\label{sec:uv_observations}

%{\bf Holmberg X-1 is surrounded by bright compact nebula \citep{Egorov2017}, whose full size in high spatial resolution Hubble Space Telescope (HST) images is $\approx2.5$\arcsec. Since the nebula contribution is much lower in UV than in the optical and near-IR ranges, short wavelengths are most preferable for studying the optical counterpart of the ULX. %Since X-1 is surrounded by bright compact ($\approx2.5$\arcsec) nebula we chose the FUV range in order to decrease flux contribution from the nebula.
%}

To prevent possible damage to the detectors from bright stars as well as to smooth image artifacts caused by bad pixels, the spacecraft is commanded to oscillate around its aiming point with amplitude of a few arcminutes and velocity of a few arcseconds per second. As a result, star positions on the detector drift with time, and one need to compensate this motion to convert a photon event list (produced in the PC mode) into an image appropriate for scientific analysis. This job can be done by various pipelines developed specially for UVIT. In this work we used JUDE\footnote{Released under the Apache License 2.0, and archived at the Astrophysics Source Code Library \citep{Murthy2016ascl}. The latest version can be downloaded from https://github.com/jaymurthy/JUDE. The manual is published in \citet{manual}.}.

The pipeline reads the FUV/NUV Level-1 data, extracts the photon events from each frame and adds them into an image tracing the spacecraft motion by bright stars (see description of the algorithm in Sec.~3.8 of \citealt{Murthy2017}). %in the corresponding VIS frame (see description of the algorithm in \citealt{Murthy2017}).
% по VIS крайне хреново исправляется траектория, поэтому использовались стэки по ~200 кадров для детектирования звезды и по нескольких десятков кадров для определения ее положения центроидом - все как описано в секции 3.8. По крайней мере, я так сделал картинку за 2 января 2020. Деталей того, что индусы делали - не знаю, у нас есть только их текст, описывающий это, который ты собственно и правил.
Each sub-observation (continuous segment between two passages through SAA) is treated independently. The images of individual segments are then astrometrically calibrated to place them all on a common reference frame and co-added. The resulting PSFs of the FUV co-added images have FWHMs of $1.4\arcsec-1.8\arcsec$ (varying from one observation to another, Table~\ref{table:FUVphot}); in the NUV filters, the resulting spatial resolution is slightly better, $1.2\arcsec-1.4\arcsec$. 
%\hide{The true color image of the large area around \obj\ from all \astrosat/F154W,N245M,N279N data on 2016 and image of the local area around \obj\ in F148W filter on 2016 September 29/30 are shown in left and up right panels of the Fig. \ref{fig:fig1} respectively.}
%\blue{The flux units of the final image from JUDE are counts/second (CPS).} % The true color image of the large area around \obj\ from all \astrosat/F154W,N245M,N279N data on 2016 is shown in left panel of the Fig. \ref{fig:fig1}.     %Unfortunately, some segments could not be reconstructed well (about 1/3). So we excluded them to not worsen the resolution of the final images. The exposure times in Table~\ref{tab:observations} account this.

%UVIT Level~1 imaging data was processed using specially written pipeline \jude\!\. \jude pipeline reads Level~1 data, extracts the photon events from each frame, corrects for the spacecraft motion (image registration) and adds into an image. It creates photon event lists and images of the sky. The individual images are then astrometrically calibrated to place them all on a common reference frame and co-added.

%{\bf Photometric calibration is done on the final co-added images from \jude, where the flux units are counts/second (CPS). }

%\AV{The resulting point spread function PSF in the FUV co-added images has FWHMs of $1.4\arcsec-1.8\arcsec$} \chk{(*** column of the Tab.}~\ref{tab:obs}). \AV{In NUV filters resulting spatial resolution is slightly better, $1.2\arcsec-1.4\arcsec$. 

Due to comparatively low spatial resolution of UVIT, the photometry of \obj\ with \astrosat\ faces with a number of difficulties. Besides the bright compact nebula mentioned above having a size of $\approx2.5$\arcsec, there are two stars 0.9\arcsec\ away from the source. This distance is much smaller than the obtained PSF size (the \obj\ environment as it is seen by \astrosat\ and HST is shown in Fig.~\ref{fig:fig1}). For this reason, the PSF photometry method frequently used in the case of moderately crowded fields where stars still resolvable, is useless here. We have carried out the aperture photometry with careful accounting for flux contribution coming from all the neighboring sources (the nebula, the two stars within the aperture as well as the more distant starts in Fig.~\ref{fig:fig1} whose PSF wings still can contaminate the observed ULX flux).

This work can be split into three steps. First of all, we determined the exact position of \obj\ performing the astrometric correction of the \astrosat\ images with the HST data, and measured the total flux in the certain aperture. All the flux measurements were done via the \texttt{APPHOT} package of IRAF. At the second step, we measured and subtracted fluxes of all the extraneous sources. At the final step, we performed the aperture corrections and converted fluxes into a single filter to be able to compare observation of different dates. For the NUV data, we decided to skip the last two steps because we have only three observations with this instrument.

%Photometry of \obj\ in \astrosat\ images involves a number of difficulties related with very crowded field. The distance from the object to the nearest two stars is less than 0.9\arcsec, which is much smaller than the PSF size. In addition, as noted above, the object is surrounded by the bright compact nebula with the size of $\approx2.5$\arcsec. A high spatial resolution image of the \obj\ surroundings from the Hubble Space Telescope (HST) UV data is shown in the bottom right panel of the Fig. \ref{fig:fig1}. The PSF photometry method frequently used in conditions of moderately crowded fields does not allow to obtain a satisfactory result in this case. Here we carried out an aperture photometry with the APPHOT package of IRAF which included several steps. First one is determining the exact position of the \obj\ in images and measuring the total flux in a given aperture. Second step is extraction of the contribution from all \AV{neighboring} \hide{background} sources to the total flux in aperture. In the final, the aperture correction was taken into account and the measured fluxes were converted into a single filter (see bellow for details). The second and third steps were done only for the FUV data to get the light curve for all 8 observation of \astrosat/UVIT.  %First, the source is not resolved with two nearby stars and a surrounding nebula (compare the \astrosat\ and HST UV images in Fig.\ref{fig:fig1}), which makes it impossible not only to measure its flux directly, but even to determine its exact position in the image. Second, the variable FSF 

\subsubsection{Raw aperture photometry}
We decide to perform the photometry in the same 3-pixel (1.23\arcsec) aperture (marked in Fig.\,\ref{fig:fig1}) for all the observation regardless the PSF size. This aperture still collects most of the source photons even in the worst images with the PSF FWHM of 1.8\arcsec, and, on the other hand, minimizes the contribution of neighboring stars. To be sure that our aperture is centered exactly at the ULX position, we carried out astrometric alignment of the \astrosat\ data to HST images where the ULX optical counterpart is clearly seen as a single point-like source. For this purpose, we have chosen the HST image\footnote{All the HST data used in this work were taken from the MAST archive https://archive.stsci.edu/} obtained on 2013 August 24, with the Wide Field Camera 3 (WFC3) in the F275W filter whose bandpass is close to those of the \astrosat\ NUV filters, and camera has a relatively large FOV of about 3\arcmin. Three single stars were used as reference sources. The resulting astrometry accuracy is better than 0.13\arcsec\ for each \astrosat\ observation.

%The optimal aperture radius for photometry is 3 pixels (1.23\arcsec), since, on the one hand, most of the source flux is concentrated inside the aperture even at FWHM(PSF) of 1.8\arcsec, on the other hand, this aperture size minimize the contribution of the bright parts of the nearby star cluster to the measured flux. To ensure the maximum contribution of \obj\ to the total flux in the aperture, we center it on the object. Exact \obj\ position was determined by relative astrometry between \astrosat\ and the HST, where the ULX optical counterpart is clearly seen as single point-like source. We chose the HST data\footnote{All HST data used in this work were taken from the MAST archive https://archive.stsci.edu/} obtained on 2013 August 24, with the Wide Field Camera 3 (WFC3) in the F275W filter whose bandpass is close to that of \astrosat\ NUV filters, and camera has relatively large field of view of about 3\arcmin. Three single stars were used as reference sources. The resulting astrometry accuracy is better than 0.13\arcsec\ for each \astrosat\ observation.

The global background that is largely associated with instrumental features of the UVIT detectors was estimated from an annular aperture with an inner and outer radii of 8\arcsec\ and 16\arcsec, respectively. The measured count rates were corrected for this background and then multiplied by the conversion factors (Table~\ref{table:filters}) to obtain fluxes in physical units (erg s$^{-1}$ cm$^{-2}$ \AA$^{-1}$). The resulting values together with their $1\sigma$ errors are given in Table\,\ref{tab:observations}.

\subsubsection{Flux contribution from neighboring sources} 
\label{sec:uv_neighbors}

The fluxes measured at the previous step are still contaminated by contributions from other sources: the nebula surrounding \obj\ and the stars {\tt st1} and {\tt st2} (Fig.~\ref{fig:fig1}) directly falling into the 1.23\arcsec\ aperture as well as from more distant stars located a few seconds away from the ULX and are likely to be members a single stellar cluster. The flux contribution from each particular sources varies between the observations depending on the PSF size, and must be carefully accounted for. To estimate these contributions in each \astrosat\ observation, we utilized the HST ACS/SBC/F165LP image (Table\,\ref{table:HSTphot}), the resolution of which was got worsen to the \astrosat\ level. The F165LP filter was chosen because its bandpass is close to those of the FUV filters, although it does not completely coincide with them. Therefore, in order to convert the contributions measured from the HST data into the \astrosat\ filters with the highest accuracy, we have taken into account the spectral energy distributions of these sources modeled involving HST photometry in other filters.

%Измеренные на предыдущем шаге потоки \obj\ "загрязнены" вкладом [are contaminated by] от соседних c ULX источников: окражующей ULX туманности, звезд st1 и st2 на Fig.~\ref{fig:fig1}, непосредественно попадающими в апертуру 1.23\arcsec\, а также от более далеких звезд, расположенных в нескольких секундах от ULX и, вероятсно, являющихся членами одного скопления. Величина вклада каждого из источников меняется от наблюдения к наблюдению в зависимости от размера PSF изображений, и требует тщательного учета. Для оценки этих вкладов в каждом наблюдии \astrosat\ мы использовали снимок HST, полученный с помощью камеры ACS/SBC в фильтре F165LP, разрешение которого загрублялось до требуемеого уровня. Данный фильтр был выбран поскольку его полоса близка к фильтрам FUV \astrosat, однако не совпадает с ними полностью. Чтобы пересчитать измеренные по данным HST вклады непосредственно в фильтры \astrosat\ с максимально возможной точностью, мы использовали модельные спектральные распределения энергии (SEDs, подробнее см. ниже)%модели Куруца для двух отдельных звезд, табличный спектр для туманности, starburst99 для интегральной фотометрии звезд, являющихся частью скопления), построенные с привлечением  фотометрии в других хаббловских фильтрах.

Since the extraneous sources have different nature (nebula vs stars), their contributions to the 1.23\arcsec\ aperture must be accounted for individually, otherwise one can not be able to model their SEDs with physical models. To estimate the individual flux contribution of each extraneous source, we removed (i.\,e. replaced with the local background counts) the \obj\ optical counterpart and all other extraneous sources except the current from the F165LP image. Then we decreased the spatial resolution of the image and measured flux in the 1.23\arcsec\ aperture centered at the \obj\ position. To decrease the resolution, we passed the image by a Gaussian filter\footnote{Actually, sum of two 2D Gaussians were used in order to fit the PSF core and wings, four parameters in total.}, the parameters of which were determined for each \astrosat\ observation individually by analyzing the PSF of four bright single stars in the UVIT field of view. This procedure was repeated 4 times: for the nebula, for the stars {\tt st1} and {\tt st2}, and for the other stars which we considered as members of a single stellar cluster (we will refer to them as the cluster).

Using this technique, we obtained that the contribution of both stars about the same; the 1.23\arcsec\ aperture gather $w_{\rm st}$ ranging from 0.468 to 0.566 for PSF sizes 1.8\arcsec\ and 1.4\arcsec, respectively, of their total (aperture corrected) fluxes in the F165LP filter. For the extended sources, namely the nebula and the cluster we measured $w_{\rm neb}$ and $w_{\rm cl}$ as ratios between the fluxes captured by the 1.23\arcsec\ aperture and the fluxes used for the SED modeling (see the next paragraph), the obtained values are  $w_{\rm neb}=0.664\div0.749$ and $w_{\rm cl}=0.022\div0.015$ (for 1.8\arcsec\ and 1.4\arcsec).

To construct the SEDs we carried out the aperture photometry on the HST drc images (cameras, filters and other details are in Table~\ref{table:HSTphot}). For the two stars, we used a 4-pix aperture (0.10\arcsec) for the ACS/SBC images and 3-pix for ACS/WFC and WFC3/UVIS (0.15\arcsec\ и 0.12\arcsec, respectively). Such small apertures were chosen to minimize the contribution from the nebula, which is especially bright in the visible band. The background was determined in annuli with an inner radius of $\approx0.25\arcsec$ and an $\approx0.15\arcsec$ width with small variations of about 0.02\arcsec\ depending on a particular camera. The aperture corrections were determined by measurements of 6--17 bright isolated stars in the large (0.5\arcsec\ for ACS and 0.4\arcsec\  for WFC3) and the small apertures. The zero points were taken from the PySynphot v2.01 package using the \texttt{effstim} commands. The final magnitudes and their errors for both stars are listed in Table~\ref{table:HSTphot}. The provided errors correspond to the $1\sigma$ confidence intervals and include the statistical errors of the flux measurement, the aperture correction errors, the stability of zero points and the stability of the filter PSF in each particular observation.

The photometry of the nearest to \obj\ region of the cluster was carried out in a 2.2\arcsec\ aperture centered at R.A.=08:19:28.29, Dec.=+70:42:19.9 (J2000.0). The background was measured in several areas around the cluster. The obtained magnitudes are presented in Table\,\ref{table:HSTphot}. The relatively large photometric errors are resulted mainly from strong variations of the background level. The magnitude of the nebula $m_{\rm neb}=18.94\pm0.05$ was determined only in the F165LP filter, in the 1.23\arcsec\ aperture centered at the ULX.

The measured fluxes of {\tt st1} and {\tt st2} were fitted with the Kurucz models from ATLAS9 \citep{CastelliKurucz2003} accounted for the interstellar extinction with $A_V = 0.23$ (using the curve by \citealt{Cardelli1989}), measured from the ratio of the Balmer lines in the nebula around \obj\ \citep{Vinokurov2013}. The metallicity was 0.1\,Z$_\odot$ \citep{Pilyugin2014}. The best agreement was obtained for the models with $\log{g}=4.0$ and the effective temperatures and photosphere radii $T_{\rm eff} = 32.2\pm1.4$\,kK, $R_{\rm ph} = 10.4\pm0.6 R_\odot$ ($\chi^2$/dof $\approx1.2$) and $T_{\rm eff} = 29.4\pm2.1$ kK, $R_{\rm ph} = 7.2\pm0.7 R_\odot$ ($\chi^2$/dof $\approx2.7$)
for the brighter ({\tt st1}) and the fainter ({\tt st2}) stars, respectively. These parameters as well as the absolute magnitudes of M$_V = -4.7\pm0.04$ and M$_V = -3.8\pm0.05$ correspond to the star types B0--O9 and B0--B1 IV-V (e.g., \citealt {Straizys1981}).

%Наблюдаемые SED обеих звезд аппроксимировались моделями Kurucz из ATLAS9 \citep{CastelliKurucz2003}, свернутыми с скривой межзвездного поглощения \cite{Cardelli1989} с $A_V = 0.23$, измеренном по отношению линий бальмеровской серии водорода туманности вокруг \obj\ \citep{Vinokurov2013}. Металличность была принята равной 0.1\,Z$_\sun$ \citep{Pilyugin2014}. Наилучшее согласие с наблюдательными данными получено для моделей с $\log{g}=4.0$ и эффективными температурами и радиусами, равными $T_{eff} = 32.2\pm1.4$ KK, $R_{ph} = 10.4\pm0.6 R_\odot$ и $T_{eff} = 29.4\pm2.1$ KK, $R_{ph} = 7.2\pm0.7 R_\odot$ соответственно для яркой (st1) и слабой звезды (st2). Значения $\chi^2/dof$ равны 1.2 (st1) и 2.7 (st2).  Полученные значения параметров, также как абсолютные звездные величины M$_V = -4.7\pm0.04$ и M$_V = -3.8\pm0.05$, соответствуют звездам B0-O9 и B0-B1 IV-V (например, \citealt{Straizys1981}). %Модельные величины потоков в FUV фильтрах \astrosat\ в единицах $10^{-17}$ erg~s$^{-1}$\,cm$^{-2}$\,\AA$^{-1}$ составляют: $F148W(st1)=7.59, F154W(st1)=7.06, F148W(st2)=3.00, F154W(st2)=2.78$. Расчетные значения потоков в фильтре F165LP ($5.34\times10^{-17}$ and $2.11\times10^{-17}$ erg~s$^{-1}$\,cm$^{-2}$\,\AA$^{-1}$ соответственно) в пределах ошибок фотометрии согласуются с наблюдаемыми величинами.

The cluster SED was fitted with the model spectra calculated in Starburst99 assuming the metallicity of 0.1\,Z$_\odot$ and ages $2.5-5$~Myr. The interstellar extinction was varied from the Galactic value $A_V=0.09$ to the value measured by the nebula lines ($A_V=0.23$). The minimal $\chi^2/{\rm dof} = 2.8$ has been reached for the cluster of 2.9 Myr and $A_V=0.22\pm0.02$ which is in a relatively good agreement with the result by \cite{Stewart2000}. Besides the spectrum of stars themselves, the starburst99 code outputs the continual spectrum of a nebula in which these stars are immersed. We adopted the desired model SED to be \texttt{stars+0.5*nebula}.

%Аппроксимация интегрального SED измеренной области скопления проводилась с помощью модельных спектров звездных скоплений возрастом $2.5-5$~Myr и металличностью около 0.1\,Z$_\sun$, рассчитанных в Starburst99. Величина межзвездного поглощения варьировалась от Галактического значения $A_V=0.09$ до измеренного по линиям туманности вокруг \obj\ $A_V=0.23$. %Величина межзвездного поглощения $A_V = 0.13$\citep{Stewart2000} была зафиксирована. 
%Минимум значения $\chi^2/dof = 2.8$ получено для модельного спектра скопления возрастом 2.9 Myr при $A_V=0.22\pm0.02$, что неплохо согласуется с оценками \citep{Stewart2000}. Некоторая неопределенность/вариативность результата связана со вкладом туманности, чей спектр был рассчитан совместно со спектром звездного скопления. %Потоки, рассчитанные на основе модели составляют $F148W(cluster)=3.09, F154W(cluster)=2.92$ и $F165LP(cluster)=2.16$ в единицах $10^{-15}$ erg~s$^{-1}$\,cm$^{-2}$\,\AA$^{-1}$.

As a model for the nebula around \obj\ was used the tabulated spectrum\footnote{File \url{pn_nebula_only_smooth.fits}, taken from \url{https://archive.stsci.edu/hlsps/reference-atlases/cdbs/etc/source/} (MAST library).} of a planetary nebula with the extinction of $A_V = 0.23$ applied. This choice is based on the fact that photoionization-dominated shells around many ULXs (and \obj\ in particular) demonstrate spectra similar to those of planetary nebulae, with a large number of high-excitation lines at high intensities, and are believed to have similar gas ionization state (see e.\,g. \citealt{Abolmasov2007}). The use of the tabulated spectrum leaves only one free parameter~--- a normalization, which was determined from the nebula flux measured above.

%Спектральное распределение энергии туманности аппроксимировалось табличным спектром планетарной туманности pn\_nebula\_only\_smooth.fits\footnote{спектр планетарной туманности (pn\_nebula\_only\_smooth) в формате fits-таблицы взят из библиотеки спектров по адресу \url{https://archive.stsci.edu/hlsps/reference-atlases/cdbs/etc/source/}}, свернутым с кривой межзвездного поглощения с фиксированным значением $A_V = 0.23$. Основанием выбора этой модели служит схожесть состояния ионизации газа и, как следствие, наблюдаемых спектров планетарных туманностей и фотоионизационно-доминированных оболочек вокруг многих ULX (включая \obj), которые демонстрируют широкий набор линий высокого возбуждения большой интенсивности (например, \citealt{Abolmasov2007}). При таком подходе свободным параметром остается нормировка модельного спектра туманности, которая определялась на основе измеренного потока туманности в фильтре F165LP. %Модельные величины потоков составляют F148W(nebula)=$1.53$, F154W(nebula)=$1.62$ and F165LP(nebula)=$1.52$ in units of $10^{-16}$ erg~s$^{-1}$\,cm$^{-2}$\,\AA$^{-1}$. \chk{Do we need model flux errors given that they are not used anywhere?} %(приведенные величины ошибок соответствуют ошибке наблюдаемого потока в фильтре F165LP).  

The SED of each of the four considered objects was multiplied by the corresponding factor $w$ determined above to derive flux in the \astrosat\ 1.23\arcsec\ aperture and then was convolved with the bandpass of the target UVIT filter. The total flux of the four sources in the 1.23\arcsec\ aperture as well as the \obj\ flux corrected for this value for each observation are presented in Table~\ref{table:FUVphot}. Eventually, the contaminating flux turned out to be weakly dependent on the PSF size because the contribution from the two stars and the nebula decreasing as the resolution get worse is partly compensated by growing contribution from the cluster.

%Суммируя величины модельных потоков соседних с \obj\ звезд, скопления и окружающей туманности с учетом доли их излучения в апертуре 1.23\arcsec\, мы рассчитали коэффициенты перехода между фильтром F165LP камеры ACS/SBC космического телескопа им. Хаббла и FUV фильтрами \astrosat\ как соотношение потоков в этих фильтрах: $F148W(neighbours)/F165LP(neighbours)=1.23-1.25$ и \\ $F154W(neighbours)/F165LP(neighbours)=1.21$. Изменение коэффициента пересчета F148W/F165LP связано с вариациями полученных PSF в наблюдениях \astrosat\ в диапазоне $\approx1.4\arcsec-1.8\arcsec$, в то время как размер PSF для данных в фильтре F154W оказался почти одинаковым (см. величины FWHM PSF в колонке \#2 Таблицы \ref{table:FUVphot}).%Величины коэффициентов оказались слабо зависящими от размера PSF на изображениях \astrosat\. %Соответствующие рассчитанные суммарные потоки в апертуре 1.23\arcsec\ всех соседних с \obj\ источников представлены в колонке \#4 Таблицы \ref{table:FUVphot}. 
%Результирующие суммарные потоки в апертуре 1.23\arcsec\ всех соседних с \obj\ источников, пересчитанные в фильтры F148W и F154W приведены в колонке \#3 Таблицы \ref{table:FUVphot}. Отметим, что как величины коэффициентов пересчета, так и величины суммарных потоков оказались слабо зависящими от размера PSF. %, что связано с увеличением вклада звезд скопления и икружающей их HII-области по мере роста FWHM PSF.
%Скорректированные за вклад соседних объектов потоки \obj\ в фильтрах F148W и F154W в апертуре 1.23\arcsec\ приведены в колонке \#4 Таблицы \ref{table:FUVphot}).

\begin{table*}
\centering
\caption{Results of the HST photometry of \obj\ and the neighboring sources}
\label{table:HSTphot}
\begin{tabular}{llccccc}
\hline 
Camera/Filter & ~~Obs. Date~~ & ~Exp.,~ & \multicolumn{4}{c}{Magnitudes in the Vegamag system } \\ \cline{4-7}
& & s & st1 & ~~st2 & ~~cluster & ~~\obj \\
%Camera/Filter & ~~Obs. Date~~ & ~Exp., s~ & st1 & ~~st2 & ~~cluster & ~~\obj \\
\hline
ACS/SBC/F165LP & ~~2006\,Nov\,27 & 600 & $20.12\pm0.08$ & ~~$21.12\pm0.13$ & ~~$16.16\pm0.03$ & ~~$18.90\pm0.05$ \\
WFC3/UVIS/F275W & ~~2013\,Aug\,24 & 2424 & $20.88\pm0.03$ & ~~$21.84\pm0.04$ & ~~$16.81\pm0.02$ & ~~$19.49\pm0.03$ \\
WFC3/UVIS/F336W & ~~2013\,Aug\,24 & 1146 & $21.44\pm0.04$ & ~~$22.42\pm0.05$ & ~~$17.23\pm0.03$ & ~~$19.91\pm0.03$ \\
WFC3/UVIS/F438W & ~~2013\,Aug\,24 & 992 & $23.02\pm0.04$ & ~~$24.06\pm0.06$ & ~~$18.68\pm0.05$ & ~~$21.58\pm0.03$ \\
ACS/WFC/F550M & ~~2006\,Jan\,28 & 1505 & $23.13\pm0.04$ & ~~$24.07\pm0.05$ & ~~$18.84\pm0.03$ & ~~$21.84\pm0.04$ \\
ACS/WFC/F814W & ~~2006\,Jan\,28 & 600 & $23.46\pm0.07$ & ~~$24.08\pm0.10$ & ~~$18.65\pm0.03$ & ~~$21.40\pm0.03$ \\
\hline
\end{tabular}\\
\end{table*}

\begin{table*}
\centering
\caption{Details of the \astrosat\ photometry. $B_{\rm neib}$~--- total flux from the neighboring sources, captured by the 1.23\arcsec\ aperture, $F_{\rm net}$~--- \obj\ flux in the 1.23\arcsec\ aperture with background and $B_{\rm neib}$ subtracted. $F_{\rm F148W}$~--- net \obj\ flux corrected to infinite aperture and converted to the same F148W filter.}
\label{table:FUVphot}
\begin{tabular}{lcccccc}
\hline 
N. & Filter & ~~FWHM of~~ & Encircled & \multicolumn{3}{c}{Flux, $10^{-16}$ erg~s$^{-1}$~cm$^{-2}$~\AA$^{-1}$}  \\
\cline{5-7}
 & & PSF, arcsec & energy & $B_{\rm neib}$ & $F_{\rm net}$ & $F_{\rm F148W}$ \\ 
%Num & Filter & ~~FWHM PSF,~~ & ~Enc. En.,~ & ~Neighbours' Flux~ & ~~\obj\ Flux~~ &\obj\ flux \\
%& & \arcsec &  & in aperture 1.23\arcsec  & & (in F148W) \\
\hline
1 & F148W & $1.40\pm0.07$ & $0.61\pm0.05$ & $2.36\pm0.07$ & $1.24\pm0.13$ & $2.03\pm0.23$\\
2 & F154W & $1.53\pm0.07$ & $0.57\pm0.05$ & $2.35\pm0.07$ & $1.25\pm0.14$ & $2.30\pm0.28$\\
3 & F154W & $1.56\pm0.07$ & $0.57\pm0.05$ & $2.35\pm0.07$ & $1.05\pm0.14$ & $1.91\pm0.27$\\
6 & F148W & $1.83\pm0.06$ & $0.51\pm0.04$ & $2.33\pm0.07$ & $1.15\pm0.11$ & $2.27\pm0.23$\\
7 & F148W & $1.77\pm0.05$ & $0.51\pm0.04$ & $2.33\pm0.07$ & $1.22\pm0.11$ & $2.39\pm0.23$\\
8 & F148W & $1.58\pm0.05$ & $0.58\pm0.03$ & $2.35\pm0.07$ & $1.24\pm0.11$ & $2.12\pm0.20$\\
9 & F148W & $1.70\pm0.09$ & $0.53\pm0.06$ & $2.34\pm0.07$ & $1.11\pm0.11$ & $2.11\pm0.24$\\
10 & F148W & $1.68\pm0.09$ & $0.52\pm0.05$ & $2.34\pm0.07$ & $1.06\pm0.11$ & $2.04\pm0.23$\\
\hline
\end{tabular}\\ 
\end{table*}

\subsubsection{Aperture corrections and flux conversion between the filters}

The final step of the FUV data reduction involved the aperture correction of the obtained net \obj\ fluxes and conversion to the same F148W filter in order to able to study the source variability. The conversion was needed for the two observations of 2016 that were carried out in the F154W filter. To do it, as in the previous section, we modeled the object SED using the HST photometry. 

%Финальный этап обработки/анализа FUV данных \astrosat\ предусматривал коррекцию за апертуру полученных "чистых" потоков \obj\ в апертуре 1.23\arcsec\ и их пересчет в фильтр F148W для последующего исследования переменности объекта. Пересчет требовался для двух из 8 наблюдений, выполненных в фильтре F154W в ноябре и декабре 2016 года, и проводился с помощью описанной в разделе \ref{neighbors} методике.
The aperture corrections for the \astrosat\ images were calculated by measuring the fluxes of four bright isolated stars in apertures of 3 (1.23\arcsec) and 50 ($\approx20\arcsec$) pixels. The background level was estimated in annuli with an inner radius and width of 75 and 25 pixels, respectively. The obtained values in a form of encircled energy and their 1$\sigma$ errors are listed in Table\,\ref{table:FUVphot}.

%Апертурные поправки на изображениях \astrosat\ вычислялись путем измерения потоков четырех ярких одиночных звезд в апертурах 3 (1.23\arcsec) и 50 ($\approx20\arcsec$) пикселей. Уровень фона оценивался в кольцах с внутренним радиусом и шириной равными 75 и 25 пикселей соответственно. Результаты измерений в терминах encircled energy вместе с 1$\sigma$ ошибками представлены в колонке \#5\ Таблицы \ref{table:FUVphot}. 
 
%Фотометрия для оптического двойника ULX на изображениях HST проводилась аналогично измерениям потоков двух ближайших к нему звезд (см. раздел 3.1.2), результаты приведены в Таблице \ref{table:HSTphot}. %Для аппроксимации измерений была выбрана модель чернотельного излучения.

The HST photometry of \obj\ was carried out using the same images and the same methods as described above for the two neighboring stars. The resulting magnitudes are listed in Table\,\ref{table:HSTphot}. 

\obj\ is a confirmed optically variable source which can potentially cause problems of two kinds. The first one is because the available HST flux measurements in different filters are not synchronous, and, therefore, they may not fit the single model. The second is that the SED, obtained from the HST data, may not be applicable at the time of our observations if the source state had changed. Fortunately, we have fount that this is not the case. Optical variability of \obj\ is not high ($\Delta m_V\approx 0.07$ in the HST observations of 2006--2007, \citealt{Tao2011}); some authors have already successfully fitted its SED \citep{Tao2012HoIIoptical,Vinokurov2017}. Fitting the fluxes from Table\,\ref{table:HSTphot} with a black body\footnote{The most red filter F814W was excluded because the variability of \obj\ is known to be the highest in the IR range.}, we obtained $T_{\rm eff}=35.5\pm2.9$~kK with $\chi^2/{\rm dof} = 3.6$ which is close to the result by \cite{Tao2012HoIIoptical}. After convolving the SED model with the bandpass curves, we derived fluxes $M_{\rm F148W}=(2.49\pm0.08)\times10^{-16}$ and $M_{\rm F154W}=(2.39\pm0.08)\times10^{-16}$ erg~s$^{-1}$\,cm$^{-2}$\,\AA$^{-1}$. These model values based on the HST photometry are close to those by \astrosat\ (Table\,\ref{table:FUVphot}), so we can conclude that the source state did not change significantly, and the problem number two does not take place here. Moreover, since the F148W and F154W filters are very close (their bandpasses intersect by $\approx80$\%, Table\,\ref{table:filters}), small variations in the SED shape can produce only negligible conversion error. The conversion was done by multiplying the net F154W fluxes by the $M_{\rm F148W}/M_{\rm F154W}$ coefficient of $1.040\pm0.010$. The final aperture corrected fluxes in the F148W filter are presented the last column of Table\,\ref{table:FUVphot}. The uncertainties correspond to $\sigma$ confidence intervals and are a square root of a sum of squares of the following components: the statistical errors of the flux measurements in the aperture 1.23", the uncertainties related the subtraction of the neighboring sources and the uncertainties of determining of the aperture corrections.

To control the obtained results, in particular, the issue that concerns the flux contribution from the neighboring sources varying with the PSF size, we repeated the reduction by another way. Before performing the photometry, we decreased the resolution of each \astrosat\ image to the worst one ($\approx1.8$\arcsec, observation \#6) which have to equalize the aperture corrections and the contributions between the observations. The images were smoothed  with a 2D Gaussian filter whose width in each direction was chosen in that way to make the aperture corrections and FWHMs of single stars the same (within error) as observed in the image \#6. Then we measured the net \obj'\ fluxes assuming the total contribution from the neighboring sources to be equal to that determined in Sec.\,3.1.2 for the observation \#6. The fluxes obtained  by both methods turned out to be in a very good agreement.

\subsection{X-ray observations}

\begin{figure*}
\centering
\includegraphics[scale=0.44]{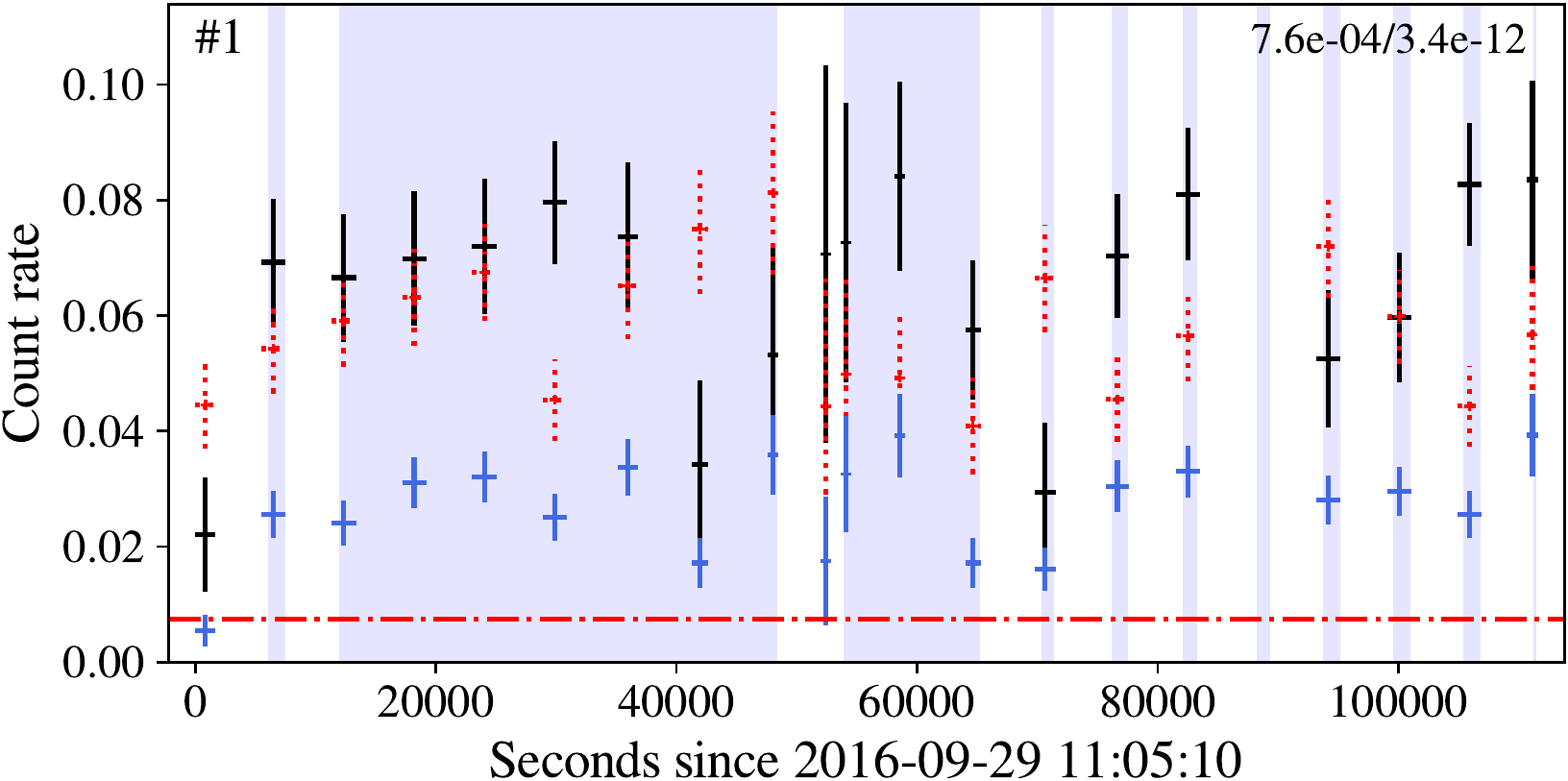}\hspace{5mm}
\includegraphics[scale=0.44]{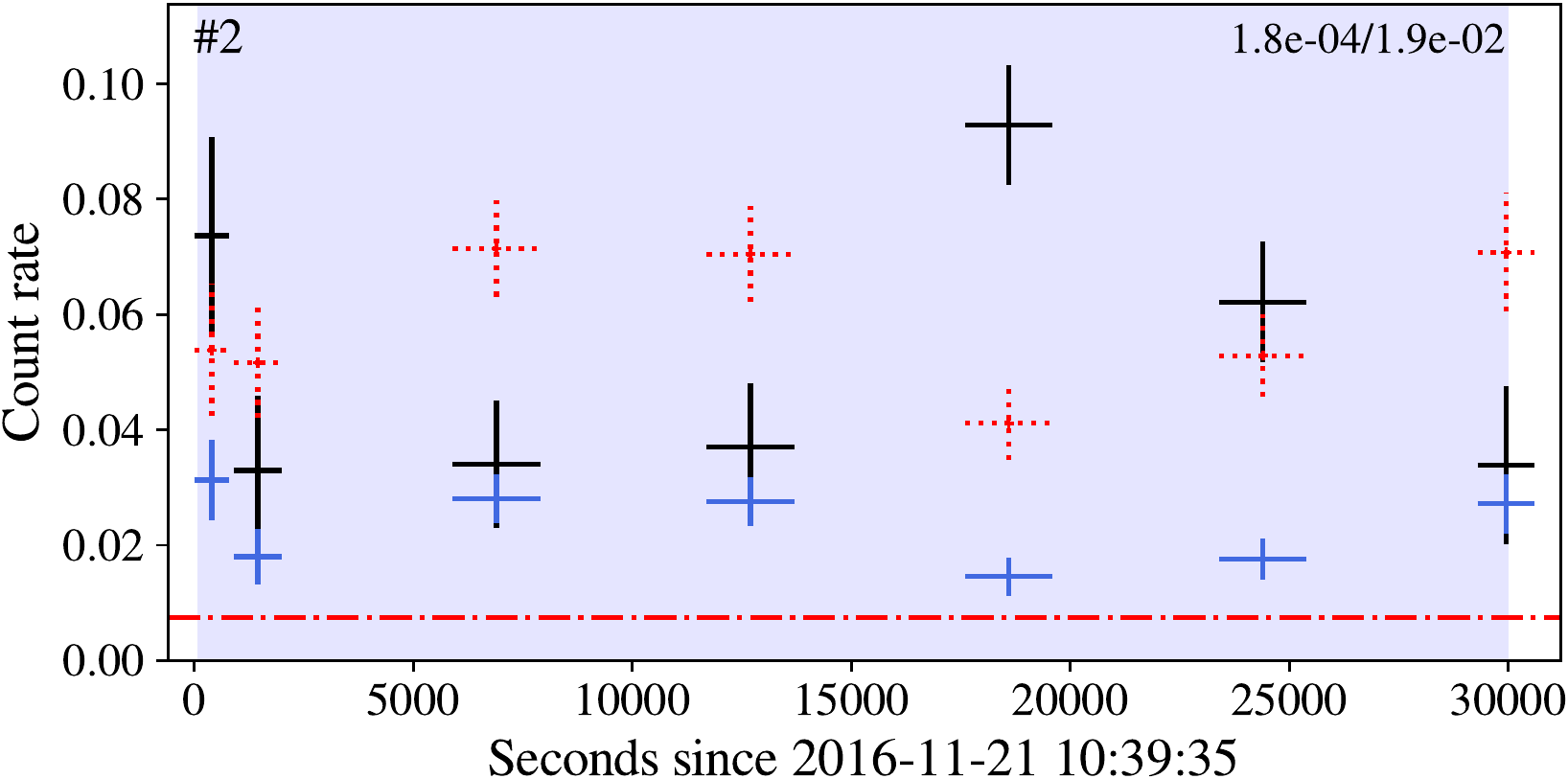}\vspace{3mm}
\includegraphics[scale=0.44]{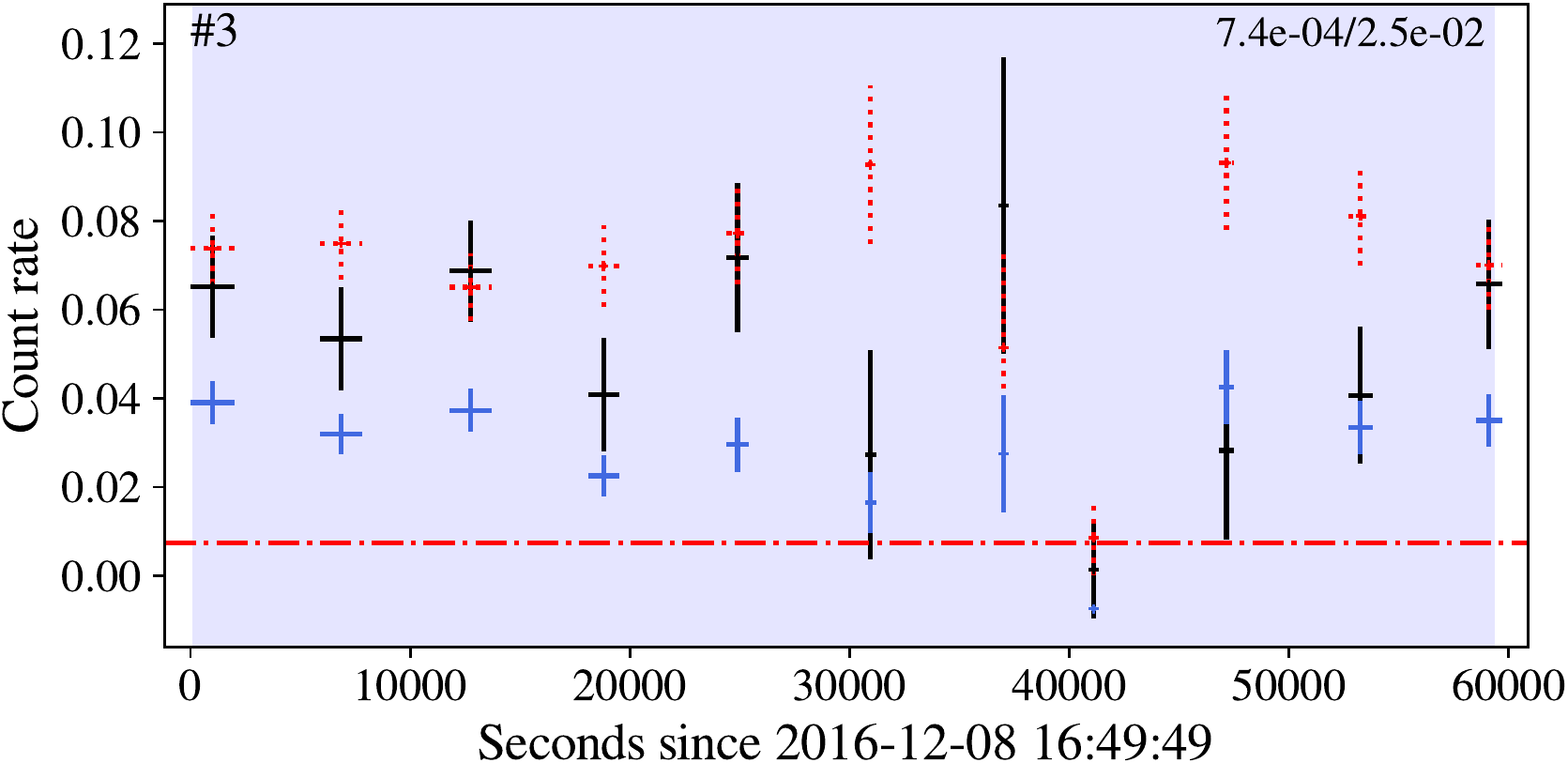}\hspace{5mm}
\includegraphics[scale=0.44]{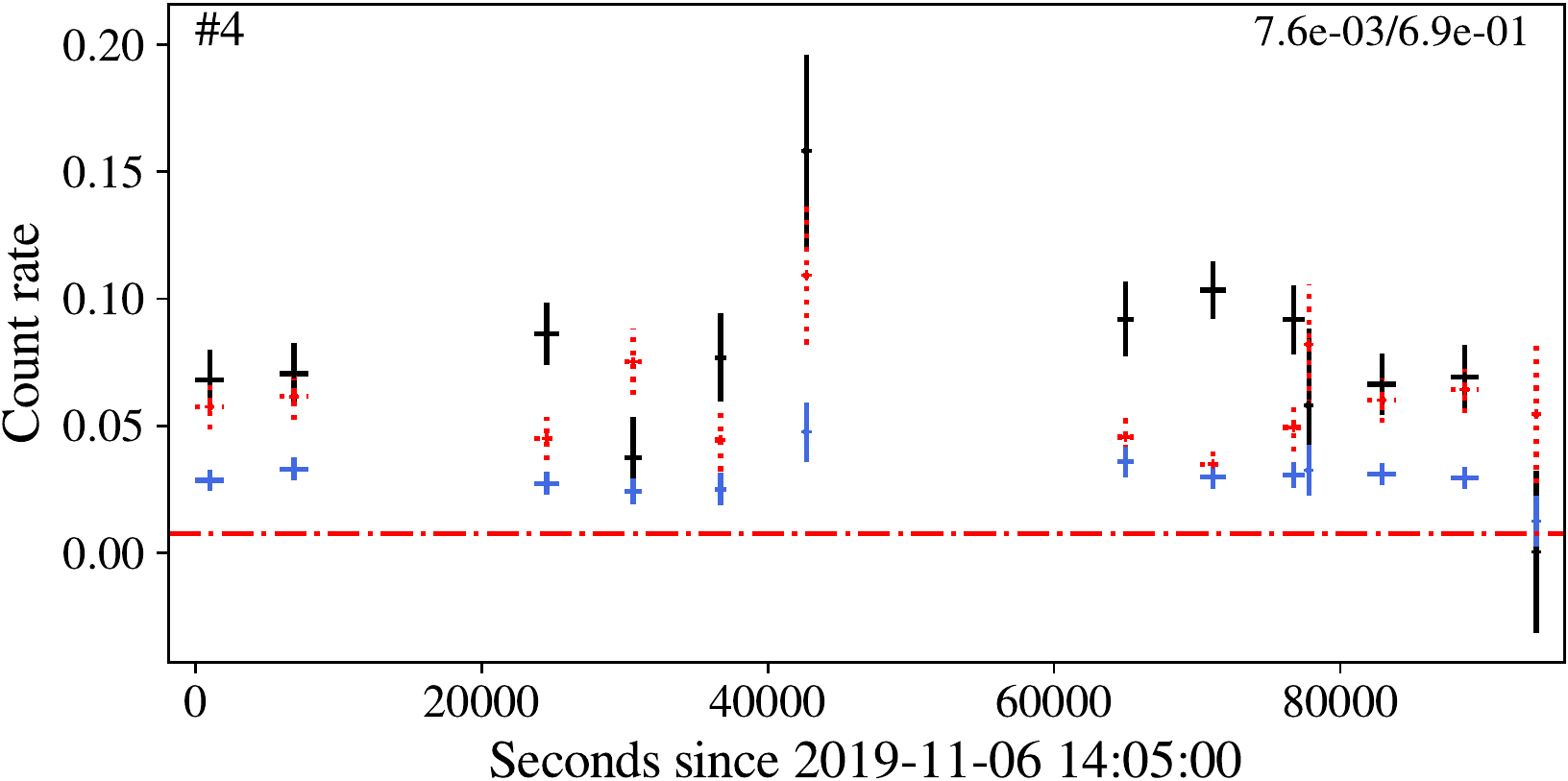}\vspace{3mm}
\includegraphics[scale=0.44]{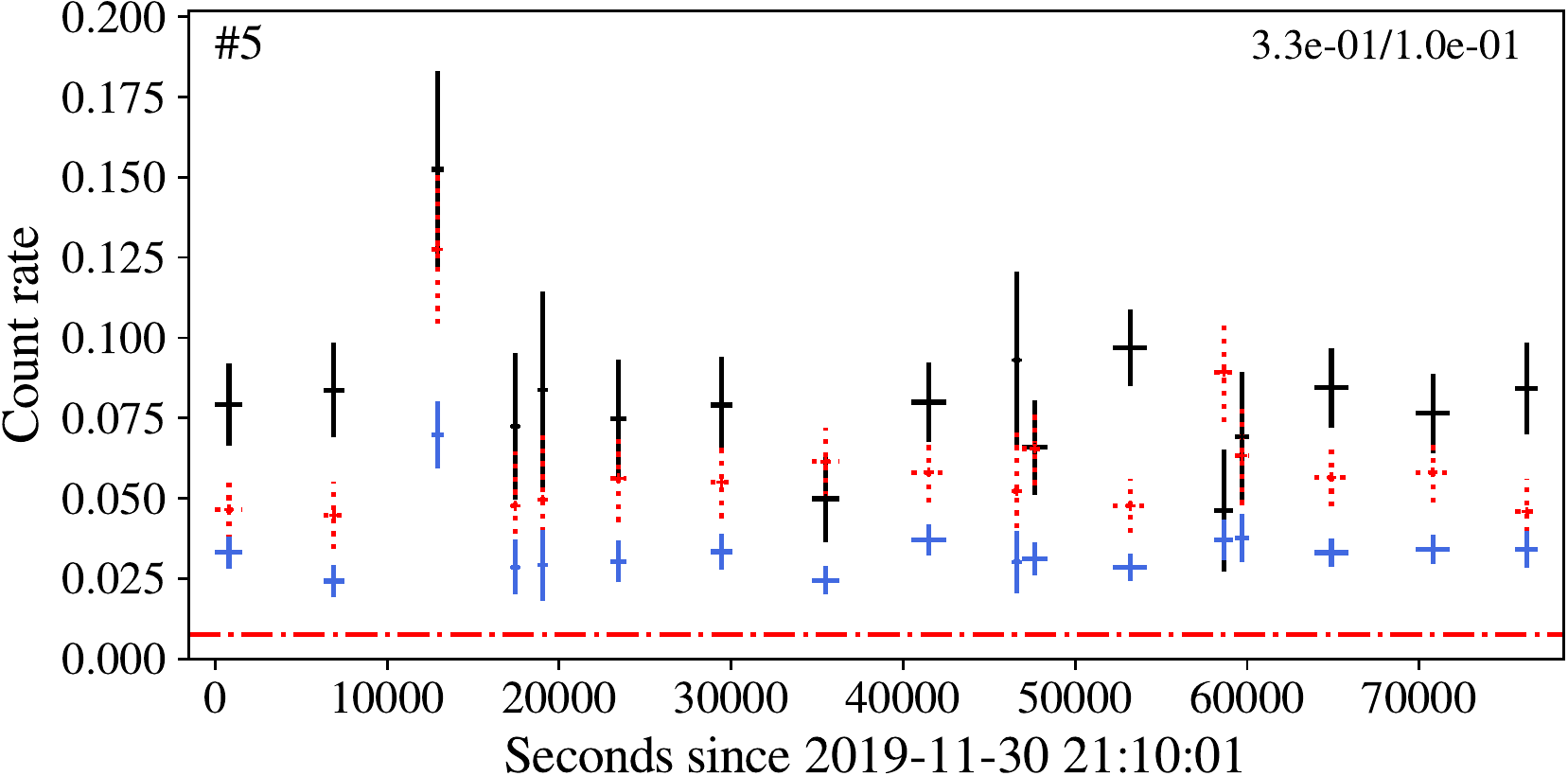}\hspace{5mm}
\includegraphics[scale=0.44]{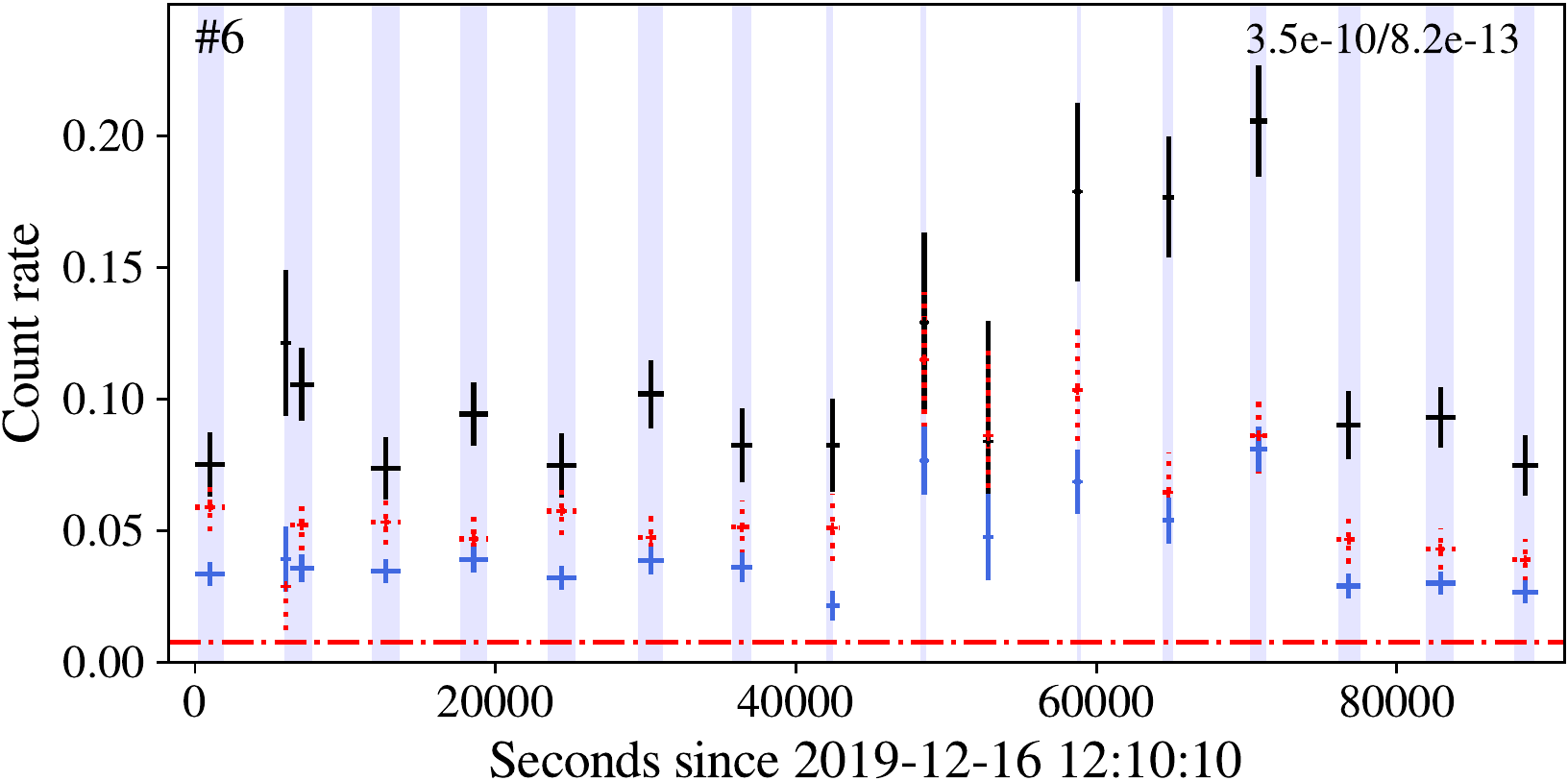}\vspace{3mm}
\includegraphics[scale=0.44]{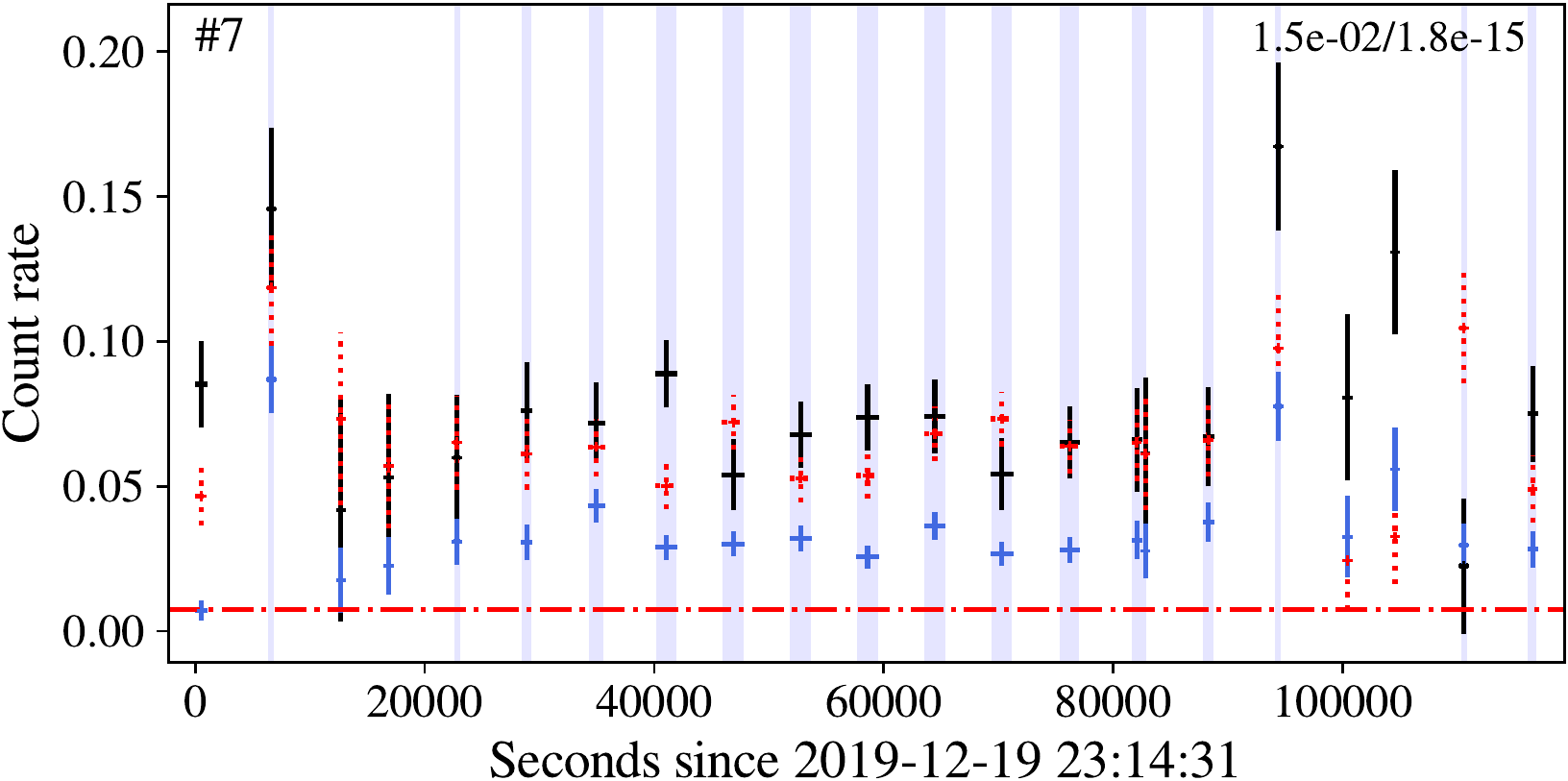}\hspace{5mm}
\includegraphics[scale=0.44]{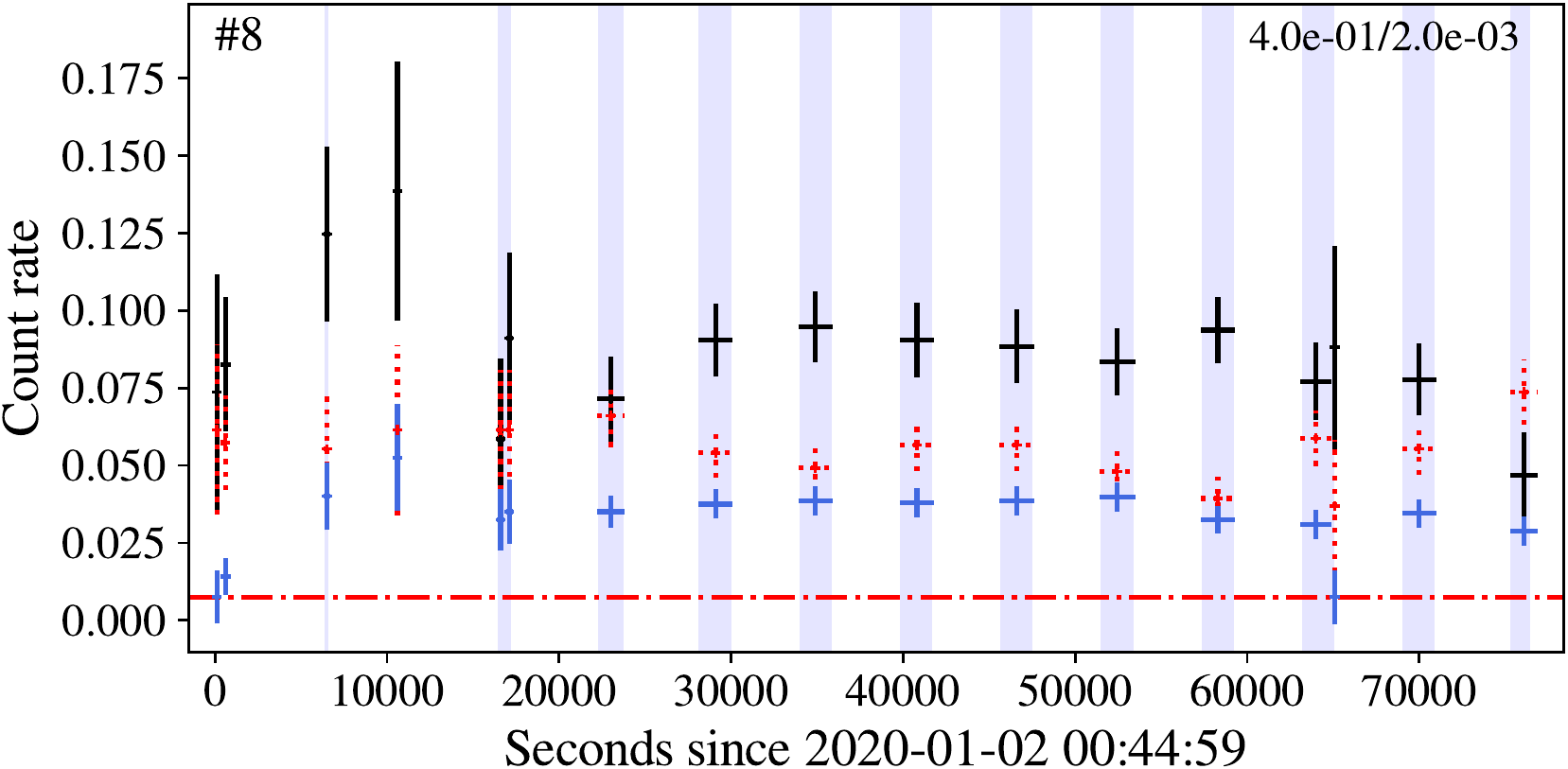}\vspace{3mm}
\includegraphics[scale=0.44]{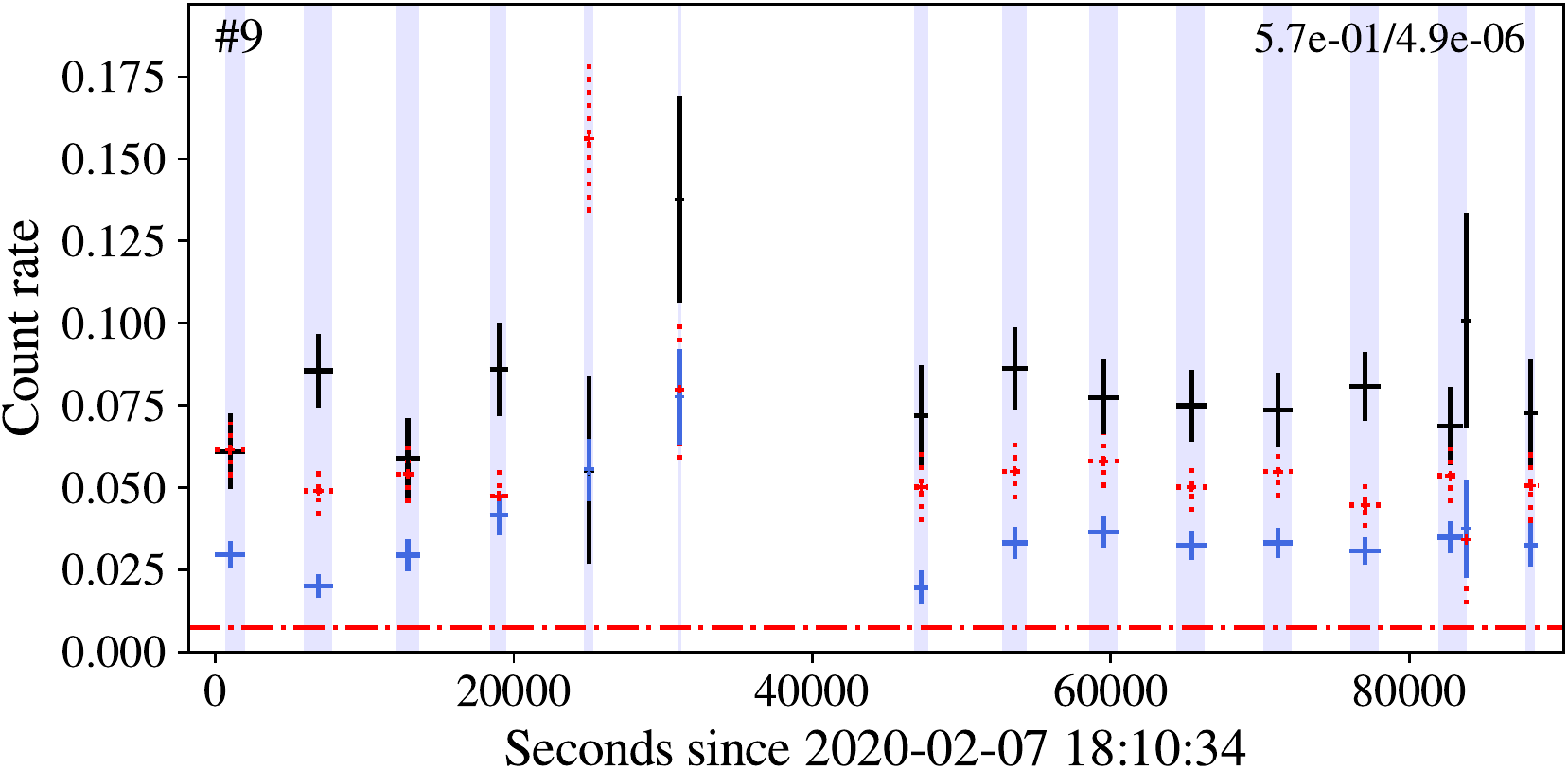}\hspace{5mm}
\includegraphics[scale=0.44]{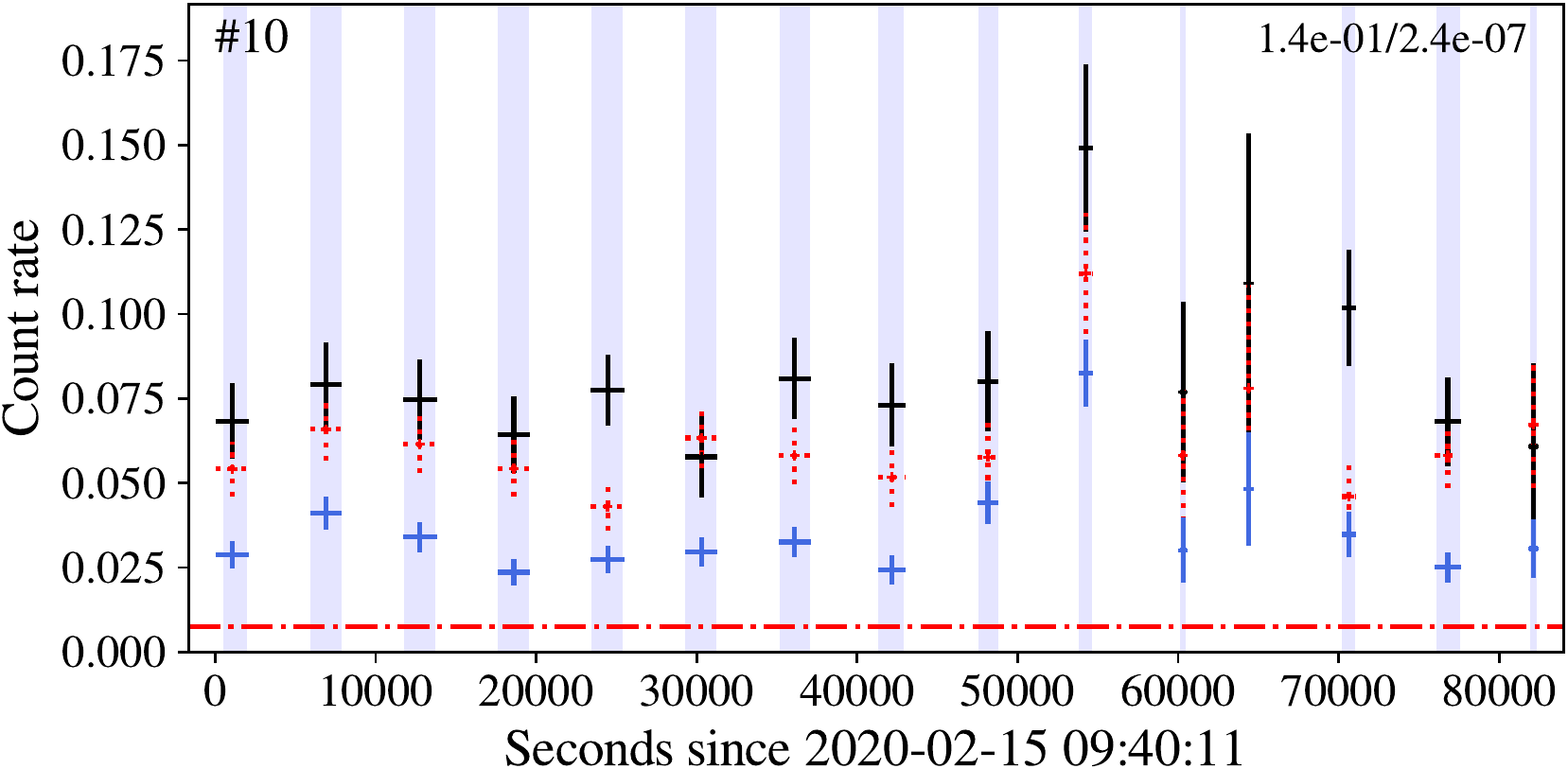}\vspace{3mm}
\caption{X-ray light curves in the 0.7-7.0\,keV range. Due to the faintness of \obj\ with respect to the \astrosat\ background, we extracted the source counts in two variants (see text): from the 5\arcmin-aperture with subtraction of the `standard' background (the same for all the observations, method I) and from the 16\arcmin-aperture with the background measured directly from the observed data (method II). The corresponding net count rates are shown by blue and black points. The background count rate re-scaled to the 16\arcmin\ source aperture is shown by red points. The red dash-dotted line denotes the level of the standard background (0.0074 cnt/s) re-scaled to the 5\arcmin-aperture. The vertical stripes show the time intervals when UVIT was turned-on. The numbers in right top corners are p-values of the variability significance for both methods.}
\label{fig:xray_lcurves}
\end{figure*}

\begin{figure*}
\centering
\includegraphics[scale=0.7]{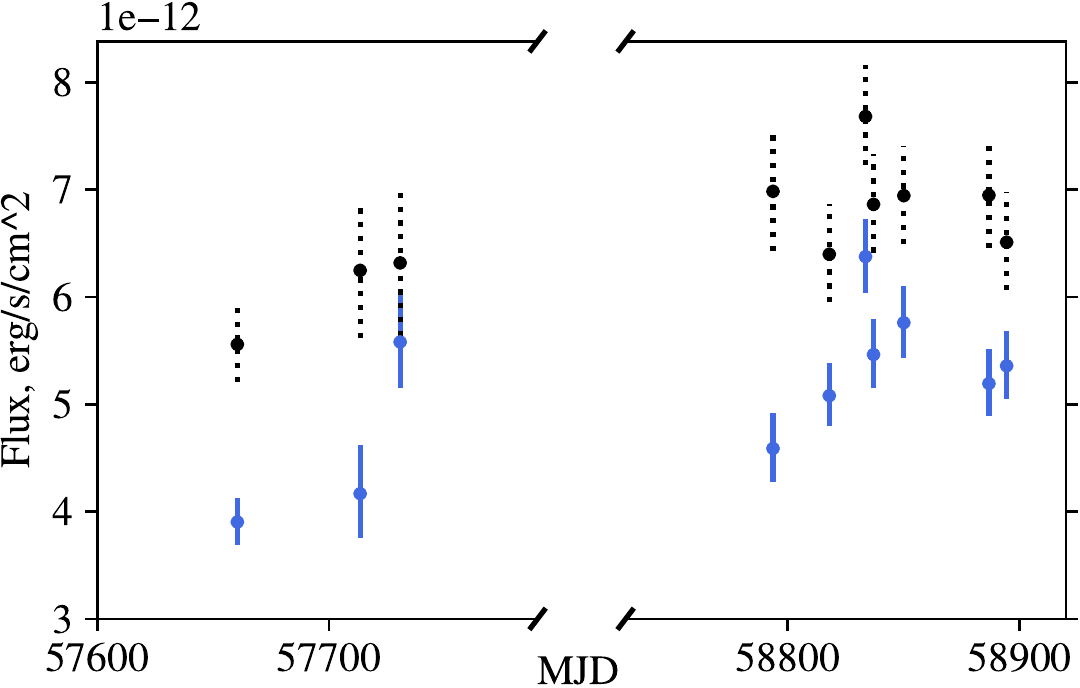}\hspace{5mm}
\includegraphics[scale=0.7]{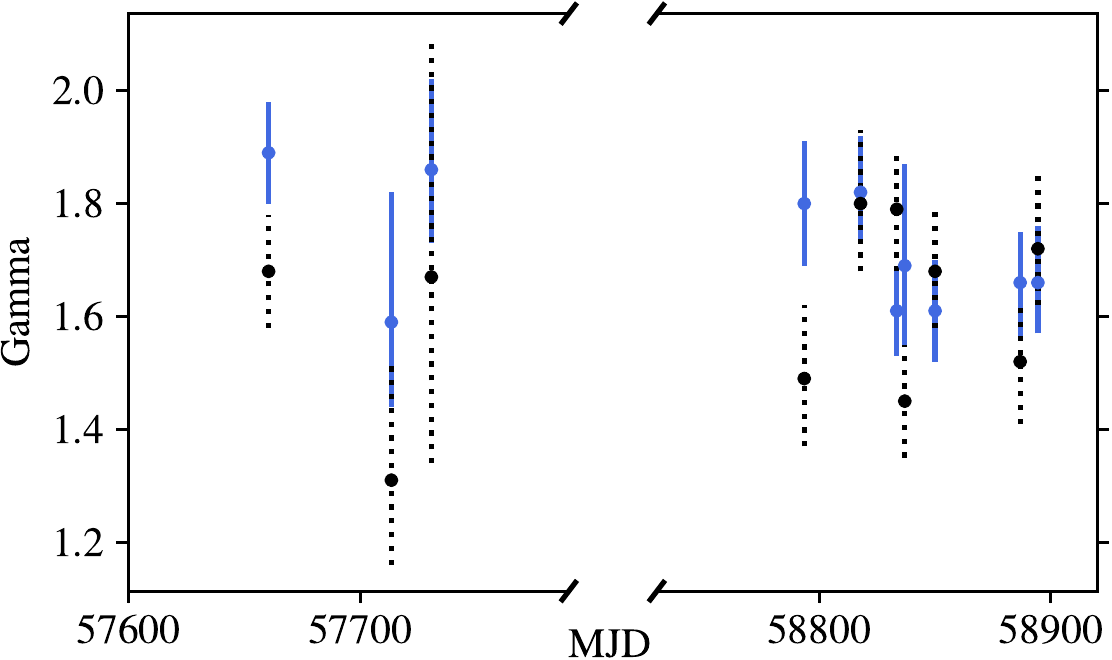}
\caption{Net X-ray flux ({\it left}) and power-law index $\Gamma$ ({\it right}). Colors, the same as in Fig.\,\ref{fig:xray_lcurves}, denote two extraction methods (see text): blue~--- method I (5\arcmin), black~--- method II (16\arcmin).}
\label{fig:xray_results}
\end{figure*}

The SXT data were processed using the official pipeline\footnote{\label{note:sxt_pipe}The software, calibration files and comments from the SXT team are available at \url{https://www.tifr.res.in/~astrosat_sxt/dataanalysis.html}}. The Level~2 files of individual segments (orbits) were merged to get the combined events files using the \texttt{sxtevtmergertool}. In order to extract light curves and spectra we used the \texttt{xselect} task distributed as a part of Heasoft\,v6.28 package. Spectral analysis was carried out with \texttt{xspec}\,v12.11.1.

For the standard analysis, the SXT team recommends\footnotemark[\value{footnote}] extracting the source counts from a circular aperture of 16\arcmin\ radius (covering about 95\% PSF) in the $0.3-7.0$\,keV range and using the background spectrum obtained by a special calibration set of observations\footnote{\texttt{SkyBkg\_comb\_EL3p\_Cl\_Rd16p0\_v01.pha}, we will refer to it as `standard background'}. This is argued by the fact that the PSF profile of SXT has a wide component extending up to 19\arcmin\ (while the FOV is 40\arcmin), and therefore the `custom' background measured in areas around the source may be contaminated by the source counts. It has to lead to overestimation of the background. However, our inspection of the \obj\ observation has shown that the custom background appeared to be lower than the standard one (rescaled to the same area) in all the cases. Moreover, we found that the background varies between the observations by up to 30\%, so the subtraction of the same level for all the data sets may yield incorrect estimates of the net fluxes, especially for such a faint source as \obj. In this regard we decided to present our analysis in two variants. The first one (hereafter method I) is the `conservative' where the source counts are extracted from a 5\arcmin-aperture and the standard background is used. This aperture has to still cover 50\% of the PSF but collect much less background counts. We found that, in this variant, the background contribution to the total flux collected by the aperture is about 20\% which make the problem of choosing  a particular background less important. In the second variant (method II), we used a 16\arcmin-aperture and custom background extracted from a number of regions in the same frame. The regions, rectangular and circular, were placed individually for each data set in a way as to cover maximal area and be outside both the 16\arcmin\ source region and aside from calibration sources at the chip corners. In this variant, the source and background contribution are about equal. Additionally, we excluded energies below 0.7~keV that are dominated by background.

In Fig.~\ref{fig:xray_lcurves} we present the net (background subtracted) light curves for both methods. Before the extraction, the event files were filtered by energy channel with expression \texttt{PI=70:700} which roughly corresponds to the energy range 0.7-7.0~keV. The count rate of the standard background in this range was obtained via \texttt{xspec} ($7.4\times 10^{-3}$~cnt/s). The light curves are binned to have one point per each continuous segment with duration of 200\,s or more. The shorted segments were dropped.

Despite high measurement errors, the source exhibit moderate variability. To assess its significance, we calculated a probability that a null-hypothesis of a constant count rate is consistent with the data using the $\chi^2_{n-1}$ distribution, where $n$ is the number of data points. The obtained p-values are shown in top right corners of each panel  of Fig.~\ref{fig:xray_lcurves}.

To obtain fluxes in physical units we carried out a spectral analysis. Due to relative low number of accumulated counts, we decided to consider only the simplest models: a singe multi-color disk or a single power law modified for interstellar absorption, and use the Cash statistic \citep{Cash1979}. The spectra were grouped with the \texttt{grppha} task to have a minimum of 1 count per bin. The response matrix (RMF) was the standard one, \texttt{sxt\_pc\_mat\_g0to12.rmf}; the ancillary responses (ARFs) were produced individually for each spectrum using the \texttt{sxt\_ARFModule\_v02.py} script with correction for vignetting enabled. This is strongly recommended when the SXT is not a primary instrument, so a source is shifted aside from the optical axis. In our observations, the shift was about 4\arcmin\ in all the observations except \#2 and \#3 where the shift was 9\arcmin\ and 11\arcmin, respectively. For these two  the vignetting correction changed  the fluxes by about 15\% and 25\% respect to uncorrected ones; in other cases the corrections were 6--9\%. The interstellar absorption was introduced by the \texttt{tbabs} model with the $N_{\rm H}$ restricted to be not lower than the Galactic value of $5.8\times 10^{20}$\,cm$^{-2}$ received via the \texttt{nh} tool of the Heasoft package. 

Both the disk and power law models have provided acceptable fits with the reduced $\chi^2_r$ from 0.8 to 1.2 ($T_{\rm in}\sim 1.2$~keV, $\Gamma \sim 1.6$ and $N_{\rm H}$ near its lower limit). Nevertheless, the power-law model gave smaller values of $\chi^2_r$ (by 5\%--15\%) in almost all observations, and we decided to derive fluxes from it. The unabsorbed source fluxes with 1-$\sigma$ errors were measured via the \texttt{cflux} convolution model, the results for both method I and II are listed in Table\,\ref{tab:observations} and shown in Fig.\,\ref{fig:xray_results}. The obtained spectral indexes are also shown in that figure.

The fluxes obtained from two apertures appeared to be different. Note that ARFs designed to account energy-depended effective area of a specific X-ray telescope when converting count rate to physical flux, being always produced for a certain aperture perform essentially an `aperture corrections' making the obtained flux independent on the aperture size. However, the fluxes from the 5\arcmin-aperture are systematically lower (but not twice, which should be without the correction for the aperture size). This may indicate that the correction is not full. On the other hand, the fluxes from the larger aperture might by overestimated because in the method II we used the background taken from peripheral areas of the FOV which may suffer from vignetting. Also, the fluxes from the smaller aperture have displayed wider scatter despite it has to collects less background counts. This might be attributed to the fact that a small aperture is more sensitive to random fluctuation of the PSF; nevertheless, the spectral indexes from the 5\arcmin\ aperture, in contrast, appeared to be more stable. So we cannot prefer any one of these two results.

\section{Results and Discussion}
\astrosat\ observed \obj\ 10 times from September 2016 to February 2020. Over all these observations, the object have demonstrated a rather weak variability in both the X-ray and UV ranges. The maximum scatter in X-rays is about 1.5 times (between \#1 and \#6). Inside the separate observations, the variability amplitude is higher. In the light curves (Fig.\,\ref{fig:xray_lcurves}), one can see flux changes with a factor of 2--3 from the mean level at time scale of $\sim 10$~ks or even shorter in five data sets (\#1, 6, 7, 9, 10). Such a behavior is known for \obj\ (\citealt{Kajava2012,Gurpide2021}). \cite{Kajava2012} note that fast variations of the X-ray flux do not lead to spectral changes. It has been proposed that apparent flux variations may be produced by relatively cold gas clouds which diminish the X-ray radiation for the observer intersecting line of sight  \citep{Middleton2015,Pinto2017}.

The averaged over all the observations and over two analysis methods source luminosity is $L_{\rm 0.7-7\,keV} \approx 8\times10^{39}$ erg~s$^{-1}$ (assuming isotropic emission and distance of 3.39 Mpc). In our observations, we found \obj\ in a hard state with $\Gamma \approx 1.6$--1.8 (Fig.\,\ref{fig:xray_results}). It is interesting that based on early XMM-Newton observation (2002-2010), \obj\ was considered as a typical soft ULX, with $\Gamma\sim 2.4$ \citep{Sutton2013,Pintore2014}. Later it became much harder (2013, $\Gamma\sim 1.9$, \citealt{Kobayashi2019,Gurpide2021}) and remained so by the time of our \astrosat\ observations. Moreover, the blue points in Fig.\,\ref{fig:xray_results} indicate hardening of the ULX to $\Gamma\sim1.6$ since December 2019. Our inspection of the XMM-Newton observation ObsID\,0843840201 taken on 23 March 2020 has shown that the source indeed exhibited a hard spectrum with $\Gamma=1.66\pm0.6$.

\begin{figure}
\centering
\includegraphics[scale=0.7]{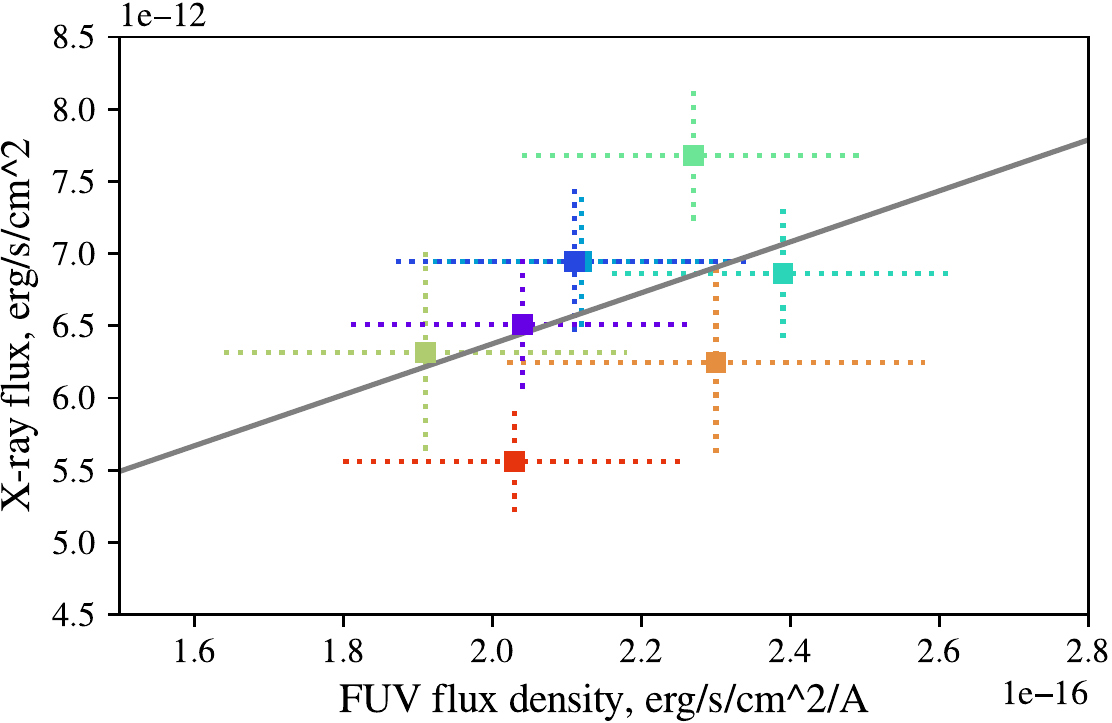}
\caption{X-ray fluxes obtained via method II (16\arcmin\ aperture) against UV flux densities at 148.1~nm (F148W filter). Colors are in the rainbow order from red (observation \#1) to violet (\#10). Formal Pearson correlation coefficient is $R=0.45\pm0.36$.}
\label{fig:xray_UV}
\end{figure}

In the UV range, the mean flux density in the F148W filter ($\approx1500$ \AA) is $F_{\rm F148W}=2.15\times10^{-16}$ erg~s$^{-1}$\,cm$^{-2}$\,\AA$^{-1}$, that correspond to luminosity of $L_{\rm F148W}=1.30\times10^{38}$ erg~s$^{-1}$ calculated using effective bandwidth of F148W filter (Table \ref{table:filters}). The variability amplitude is about 25\% of the minimum value which only slightly exceeds the 2-$\sigma$ errors of the individual measurements. The $\chi^2$ test against the null-hypothesis of constant flux yields a p-value of $\approx 0.89$, it means that the detected variability is not significant, and we can only speak about its upper limit. The weakness of the \obj\ UV variability is also evidenced by the HST measurements: the flux obtained from the SED modeling (Sec.\,3.1.3) and converted to the \astrosat\ F148W filter gives the luminosity of $(1.51\pm0.05)\times10^{38}$ erg~s$^{-1}$ which %$F_{\rm HST}= (2.49\pm0.08)\times10^{-16}$ erg~s$^{-1}$\,cm$^{-2}$\,\AA$^{-1}$ 
is close to the \astrosat\ values despite these observations are years apart.

%In Fig.\,\ref{fig:xray_UV} we plot the UV fluxes against the X-ray ones. Due to high uncertainties in both the X-ray and UV bands, one cannot see here any consistency between the data points. 

In Fig.\,\ref{fig:xray_UV} we plot the X-ray fluxes (only for method II because they have lesser scatter) against the UV flux densities at $\approx 1500$\AA. Due to high uncertainties in both the X-ray and UV bands, the relationship between the data points is not visible, nevertheless, the correlation between these two bands caused by heating of different gas structures by X-ray quanta is predicted by many ULX models. The X-ray radiation has to come from areas in close vicinity to the accretor. It may be inner parts of the supercritical accretion disk or/and a hot optically thick gas envelope if the accretor is a highly magnetized neutron star \citep{Walton2018puls}. The heating may affect an optically thick wind coming from the supercritical disk \citep{Poutanen2007}, distant parts from the disk itself or the donor star. The presence of the wind emitting in the UV/optical range is proven by both the 2D MHD simulations (e.\,g. \citealt{Kawashima2012}) and the observed optical spectra of ULXs \citep{Fabrika2015} which require the wind in order to describe a plenty of emission lines typical for them \citep{Kostenkov2020a,Kostenkov2020b}. Below all three options will be considered.

\begin{figure*}
\centering
\includegraphics[scale=0.75]{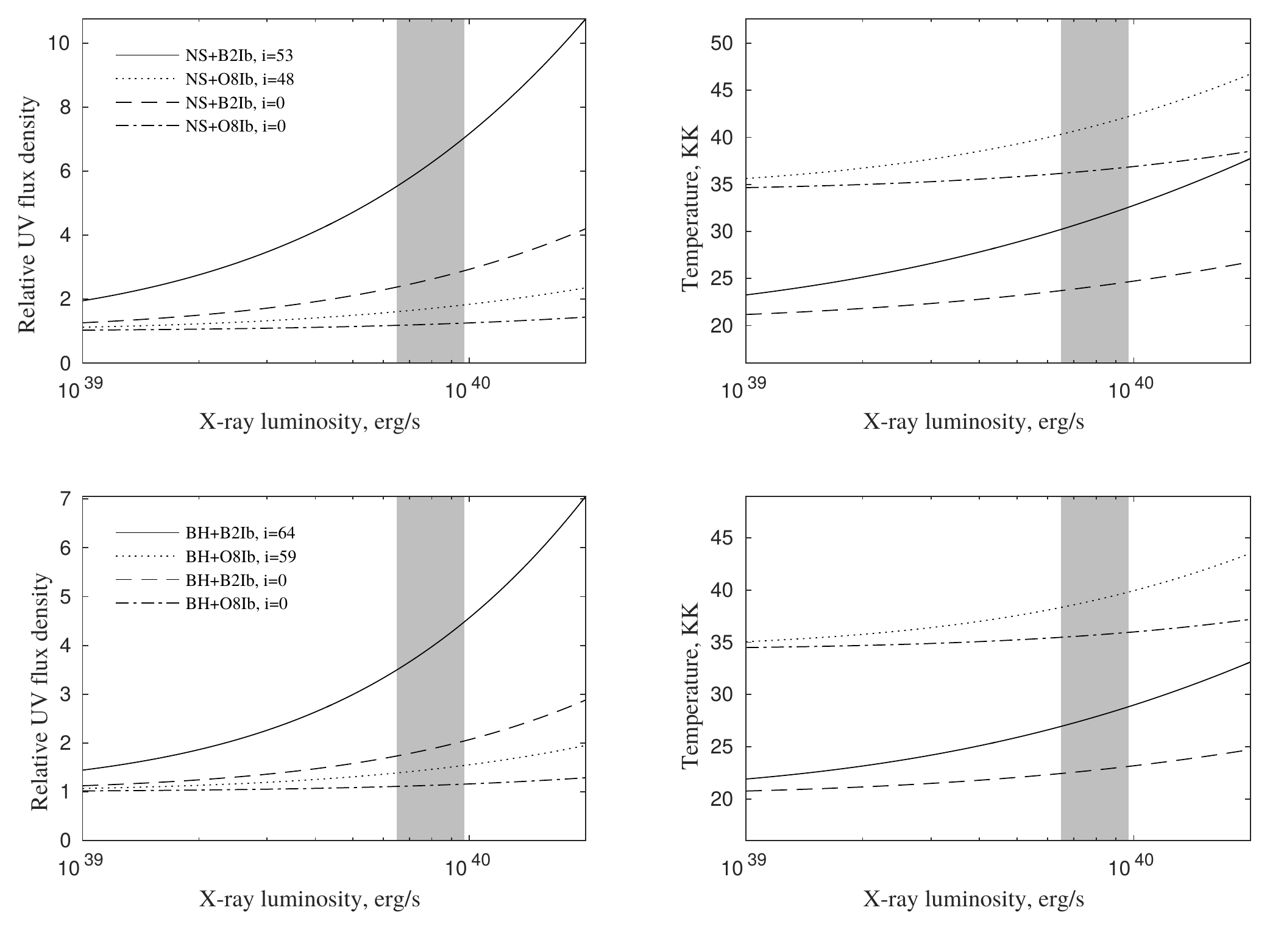}
\caption{The ratio of the flux densities at 1500\AA\ from the heated and not heated donors ({\it left panels}), and the temperatures averaged over the visible star disk ({\it right panels}) as functions of the X-ray luminosity of the accretor. Two type of the accretors: a neutron star ({\it top}) and a black hole ({\it bottom}) are considered. The donor is either a O8 or B2 supergiant. The orbit inclination either $i=0$ (view along the axis) or the maximum possible angle at which the particular system remains non-eclipsing in X-rays. Grey areas show the range of \obj\ X-ray luminosities in the \astrosat\ observations.}
\label{fig:donor}
\end{figure*}

Even if the initial accretion rate in a disk is $\dot{M}_0 \gg \dot{M}_{\rm Edd}$, its far regions that emit in the UV/optical band still release not much energy and, therefore, are supposed to be very similar to those of the standard,  subcritical ones \citep{Poutanen2007}. It has been shown that for the subcritical geometrically thin disks around low-mass X-ray binaries, the UV-optical variability induced by heating should be $F_{\rm UV} \propto L^{\alpha}_{\rm X}$ with $\alpha\sim 0.5$ \citep{vanParadijs1994,Gierlinski2009}. According to this, the X-ray flux changes by 1.5 times found by \astrosat\ should correspond to the UV variability of 20–30\%. This is close to that we have observed.

As it was shown by \cite{Fabrika2015}, the optical and IR emission of the supercritical disk wind should have a steeper dependence on the X-ray luminosity. It is due to luminosity in these ranges, being related to the Rayleigh-Jeans regions of the wind spectra, is determined mainly by the size of the emitting structure which is not constant as in the case of the standard disk but depends roughly linearly on the initial accretion rate $\dot{M}_0$. The higher the accretion rate, the wind stronger and its photosphere larger and colder; the maximum of the emitted spectrum is then shifted towards longer wavelengths \citep{Fabrika2021review}. The X-ray luminosity, in turn, depends logarithmically on  $\dot{M}_0$ \citep{ShakuraSunyaev1973,Poutanen2007}. Nevertheless, the size of the wind region that emits in the UV range remains almost unchanged because the increase of $\dot{M}_0$ just creates new colder wind parts beyond this region. Thus we can conclude that the supercritical disk wind should demonstrate the similar dependence $F_{\rm UV} \propto L^{0.5}_{\rm X}$. 

To assess UV flux variations due to heating of the donor, we carried out a modeling with following assumptions. Since the result have to strongly depend on the initial temperature $T_0$ of the star surface, we have considered two stars: B2\,Ib and O8\,Ib, that were proposed by \cite{Tao2012HoIIoptical} as possible donors of \obj. The accretor was either a black hole (BH) of 10M$_\odot$ or a neutron star (NS) of 1.5M$_\odot$. The source of X-ray quanta was assumed to be point-like, the donor fills its Roche lobe the size of which was computed via the relations by \cite{Eggleton1983}. The masses, effective temperatures and radii ($20.0\,M_\odot$, $20.3$\,kK, $23.2\,R_\odot$ and $52.5\,M_\odot$, $34.3$\,kK, $20.4\,R_\odot$ for the B2\,Ib and O8\,Ib star, respectively) were taken from \cite{Straizys1981}. The orbit of the system was assumed to be circular with a separation determined from the condition of Roche lobe filling. The observed flux density from $i$-th element of the star surface was computed using the Planck function with a temperature found from equation: $$\sigma T^4_{{\rm new},i}= \sigma T_0^4+(1-\epsilon)\cos\beta_i F_i$$ where $F_i$~--- irradiating flux, $\beta_i$~--- the angle between the directions to the X-ray source and the surface element normal, the albedo $\epsilon$ was assumed to be 0.5 \citep{Zhang1986}. In Fig.\,\ref{fig:donor} we show the new temperatures averaged over the visible disk of the star and the total flux densities at 1500\AA\ (very close to $\lambda_c$ of the F148W filter, Table~\ref{table:filters}) normalized to the values not modified by heating as functions of the X-ray luminosity. Besides two types of the donor and two types of the accretor we also considered two orientations: with the orbit inclination $i=0$ and $i=i_{\rm max}$ that was calculated individually for each case  (for a particular star size and binary separation, see the figure legend) to satisfy the condition of the system being non-eclipsing in X-rays. In the later case the donor was behind the X-ray source to make the observed effect as large as possible; the option $i=0$ was considered because majority of ULX models predicts that such systems must be viewed nearly along its axis to prevent covering of inner parts of the accretion disk by gas flows.

As it is seen from Fig.\,\ref{fig:donor}, the systems with a neutron star should yield more prominent increase of the UV flux due to their compactness. The maximal UV flux variations of $\approx27$--28\% (for the X-ray luminosities from $6.5\times10^{39}$ to $9.7\times10^{39}$ erg\,s$^{-1}$ found by \astrosat) occurs with the B2\,Ib start regardless the accretor type. In the other cases, the variations is less 20\%, with a minimum $\approx 4$\% shown by the BH+O8\,Ib system viewed from its pole. At the same time, the heated B2\,Ib start gives correct temperatures (i.\,e. close to $T_{\rm eff} = 35.5\pm2.9$\,kK obtained from the SED modeling, Sec.\,3.1.3) only with a NS viewed at $i_{\rm max}$ (maximal heating). The O8\,Ib star, in contrast, gives the correct temperature in all the cases except `NS with $i_{\rm max}$'. Thus, the observed SED temperature and low variability amplitude in the UV range make the hotter donor more preferable. Nevertheless, we should to note that we did not consider the orbital variations in this modeling. This effects, however, can be important only for systems seen nearly along the orbital plane which is thought to be unlikely for ULXs. Also, the observed variations of the X-ray flux may be caused by effects of shielding of the X-ray source by cold opaque clouds/gas flows without changing the true luminosity and, hence, the irradiating flux as well. This effects should reduce the UV variability predicted by the models.

In light of the above, one can see that the UV variability amplitudes predicted by all three considered models do not contradict the observed value due to its large uncertainties. Therefore, we have estimated the minimal required variability level in both ranges that would allow to distinguish the models. We assumed that variability could by considered as significant when at least one of data points bounces $3\sigma$ up and one $3\sigma$ down from the averaged flux. Since the UVIT uncertainties is about 10\%, the UV variability must be at least 60\% to be considered as firmly detected. For the disk/wind heating models this correspond to the X-ray variations with a factor of $\gtrsim2.5$. So, if one detects a variability of this level in X-rays but none in the UV band, it could be concluded the UV radiation is dominated by a heated O-type donor because this star have to provide the UV variability of $\lesssim30$\% regardless the accretor type and orbit inclination.

\section{Conclusion}

The Indian space satellite \astrosat\ observed \obj\ in ten epochs, eight of them simultaneously in the X-ray and UV bands. The \astrosat\ payload is similar to that of Swift which also have an UV/optical telescope but the spatial resolution of the \astrosat/UVIT is twice better. This allows to study optical counterparts of the X-ray sources residing in crowded stellar fields. 

% Хо2 является одной из лучших целей среди бона фиде УЛХ для исследования связи рентгеновского и оптического излучения... Это связано с высокой относительно подавляющего большинства других ULX яркостью объекта в оптическом и УФ диапазонах, сравнительно малым полным поглощением на луче зрения 0.23, и высокой переменностью в рентгене (до 13 раз). Измерение его потока осложняется главным образом рядом близкорасположенных звезд и яркой туманностью вокруг Х-1, яркость которой сильно падает в УФ диапазоне. Тем не менее, как мы показали в нашей работе, вклад сторонних источников может быть учтен при использовании данных HST, позволяя измерять "чистые" потоки Х-1 с точностью около 10\%. 

Though \obj\ is known as one of the most variable sources among the bona fide ULXs (with a factor up to 13, \citealt{Kajava2012}), we were unlucky to catch it in a state of low variability. We found only 1.5 times in X-rays and about 25\% (upper limit) in the UV band, which did not allow us to detect tight correlation between this bands predicted by different models. We have considered three models of heating: the heated thin disk, the wind or the O-B supergiant donor star, but cannot reject any of them with the observed variability level. To distinguish the models, we estimate the required X-ray variability as $\sim 2.5$ or higher. Our short glance on recent Swift observations have shown that \obj\ returned to the state of high variability since February 2021. This gives hope that further \astrosat\ observations will finally detect the correlation and will help to clarify the nature of the \obj\ UV-optical emission.

\section{Acknowledgments}

This research is based on observations made with the NASA/ESA Hubble Space Telescope obtained from the Space Telescope Science Institute, which is operated by the Association of Universities for Research in Astronomy, Inc., under NASA contract NAS 5–26555. These observations are associated with program IDs 10522, 10814, and 13364.

\section{Funding}

This research was supported by the Russian Science Foundation (project no. 21-72-10167 ULXs: wind and donors).

\bibliographystyle{aspb1.bst} %\bibliography{bibtexbase.bib}

\label{lastpage}
\end{document}